\documentclass[12pt,eadjoint tfnpsf]{article}

\catcode`\@=11
\@addtoreset{equation}{section}

\global\arraycolsep=1pt

\usepackage{amsbsy,amssymb,latexsym,amsfonts, amsmath}
\usepackage{mathrsfs}
\usepackage{graphicx}

\RequirePackage[dvips,usenames]{color}
\definecolor{fireblick}{rgb}{0.698039,0.133333,0.133333}

\newcommand{\beq}{\begin{equation}}
\newcommand{\eeq}{\end{equation}}
\newcommand{\bea}{\begin{eqnarray}}
\newcommand{\eea}{\end{eqnarray}}

\newcommand{\CA}{{\mathcal A}}
\newcommand{\CB}{{\mathcal B}}
\newcommand{\CC}{{\mathcal C}}

\newcommand{\CF}{{\mathcal F}}
\newcommand{\CL}{{\mathcal L}}
\newcommand{\CM}{{\mathcal M}}
\newcommand{\CN}{{\mathcal N}}
\newcommand{\CO}{{\mathcal O}}

\newcommand{\CQ}{{\mathcal Q}}

\newcommand{\CW}{{\mathcal W}}

\renewcommand\Im{{\mathrm{Im}}}

\def\Tr{\mathop{\rm Tr}}
\newcommand\tr{\mathrm{tr}}

\newcommand\diag{\mathrm{diag}}

\renewcommand{\thefootnote}{\fnsymbol{footnote}}

\begin{document}

%
%
\begin{titlepage}

\begin{flushright}
\normalsize
~~~~
SISSA  65/2011/EP-FM\\
\end{flushright}

\vspace{80pt}

\begin{center}
{\Huge Wild Quiver Gauge Theories}
\end{center}

\vspace{25pt}

\begin{center}
{
Giulio Bonelli${}^{\heartsuit, \clubsuit}$, Kazunobu Maruyoshi${}^{\heartsuit}$ 
and Alessandro Tanzini${}^{\heartsuit}$
}\\
%
\vspace{20pt}
%
${}^\heartsuit$ 
{\it International School of Advanced Studies (SISSA) \\via Bonomea 265, 34136 Trieste, Italy 
and INFN, Sezione di Trieste }\\
\vspace{7pt}
${}^\clubsuit$ 
{\it I.C.T.P. -- Strada Costiera 11, 34014 Trieste, Italy}\\
\end{center}
%
\vspace{20pt}
\begin{center}
Abstract\\
\end{center}
We study  $\CN=2$ supersymmetric $SU(2)$ gauge theories coupled to non-Lagrangian superconformal field theories
induced by compactifying the six dimensional $A_1$ (2,0) theory on Riemann surfaces with irregular punctures.
These are naturally associated to Hitchin systems with wild ramification
whose spectral curves provide the relevant Seiberg-Witten geometries.
We propose that the prepotential of these gauge theories 
on the $\Omega$-background can be obtained from the 
corresponding irregular conformal blocks on the Riemann surfaces via 
a generalization of the coherent state construction to the case of higher order singularities.


\vfill

\setcounter{footnote}{0}
\renewcommand{\thefootnote}{\arabic{footnote}}

\end{titlepage}

\section{Introduction}
\label{sec:intro}
  In the last two years an intriguing relation between non-rational conformal field theories in two dimensions 
  and ${\cal N}=2$ supersymmetric gauge theories in four dimensions \cite{AGT}
  has been under scrutiny by a variously composed community of physicists and mathematicians.
  This correspondence can be obtained by considering the $A_{N-1}$ $(2,0)$ theory on
  $\mathbb{R}^4 \times {\mathcal C}$ --
  where ${\mathcal C}$ is a punctured Riemann surface -- and a related Hitchin integrable system \cite{GMN, BT}.
  
  In the context of this correspondence, strongly coupled Argyres-Douglas points \cite{AD, MN, AS} 
  and their generalizations correspond to specific singular geometries of ${\mathcal C}$ 
  as already realized in \cite{Gaiotto}.
  A complementary picture was developed in \cite{CV} 
  in terms of quiver representations related to particular triangulations of the curve.
  
  Most of the analysis of the correspondence put forward so far had to do 
  with the gauge theory in a weakly coupled regime 
  where a Lagrangian frame can be found and used to extract relevant physical quantities,
  including the instanton partition function \cite{Nek}, 
  to compare with the CFT prediction in the limit of degenerate complex structure of the corresponding Riemann surface.
 
  In this paper we will use the CFT approach in order to access a class of $SU(2)$ gauge theories
  coupled to superconformal field theories (SCFTs) without Lagrangian descriptions,
  obtained from the compactification of the $A_1$ $(2,0)$ theory on a Riemann surface with irregular singularities.
  The Nekrasov partition function can be defined \cite{Nek} 
  as the small radius limit of a twisted index of a five dimensional theory on the circle.
  In order to compute it in the gauge theory, one should have a definition of the proper Hilbert space and Hamiltonian 
  or a Lagrangian description of the theory.
  For strongly coupled matter sectors, as for the SCFTs under consideration here, this direct approach looks hard. 
  Instead, we implement an extension of the Seiberg-Witten geometric construction to compute the solution. 
  This extension is provided by the AGT correspondence.
  We conjecture that the relevant full prepotential in the $\Omega$-background
  can be obtained by constructing the generalized coherent states 
  in the Verma module of the Virasoro algebra corresponding to the SCFT sectors. 
  
  In section \ref{sec:gauge}, we construct $SU(2)$ quiver gauge theory coupled to SCFTs.
  The basic building block, called $D_n$ theory \cite{CV}, is a deformation 
  of the $D$-type superconformal fixed point in the classification of \cite{EHIY}.
  This theory has global $SU(2)$ flavor symmetry 
  and, by gauging it, we get the $SU(2)$ gauge theory coupled to the $D_n$ SCFT.
  The Seiberg-Witten curve of the $D_n$ theory can be described 
  as a double cover of a sphere with an irregular and a regular puncture, 
  which extends Gaiotto's construction in \cite{Gaiotto} for the $A_1$ case.
  To obtain the generic $SU(2)$ quiver coupled to $D_n$ SCFTs
  one further building block is needed, namely the $T_2$ theory of \cite{Gaiotto} which consists of
  four free hypermultiplets and is associated to the sphere with three regular punctures.
  We call the resulting theories $SU(2)$ wild quiver gauge theories 
  because of the link with Hitchin systems with wild ramification that will be explained below.
  
  In section \ref{sec:geom}, we relate these gauge theories with the Hitchin integrable system.
  The $D_n$ theory corresponds to the Hitchin system with wild ramification:
  we describe how the most general irregular singularity encodes the relevant deformation parameters 
  from the corresponding superconformal fixed point 
  and provide a precise prescription defining the moduli of the theory, i.e. the vevs of relevant operators.
  
  In section \ref{sec:CFT}, we consider the CFT approach to the wild quiver gauge theory.
  We first find that the CFT counterpart of the $D_n$ theory is
  a generalization of the coherent state \cite{Gaiottostate} in the Verma module.
  These states correspond to the operators creating irregular singularities 
  of the stress-energy tensor on the Riemann surface.
  Therefore, we propose that the corresponding irregular conformal blocks describe the partition functions 
  of the wild quiver gauge theories.
  We provide evidence of this proposal by reproducing the correct
  prepotential of $SU(2)$ gauge theories coupled to one or two $D_n$ SCFTs.
  By studying the insertions of degenerate fields on the irregular conformal blocks
  we then obtain the quantization of the $SL(2,{\mathbb C})$ Hitchin system with wild ramification.

  We conclude with several open questions in section \ref{sec:conclusions}.
  In appendix \ref{sec:period}, we show the calculation of the integral of the Seiberg-Witten differential
  which is used in the comparison with the CFT approach.

\section{Gauge theories with strongly coupled sectors}
\label{sec:gauge}
  In \cite{CV} (see also \cite{CNV}), a class of $\CN=2$ gauge theories 
  associated with a Riemann surface with higher order singularities was discussed.
  Such a class of theories is specified by BPS quiver diagrams related with triangulations of the Riemann surface.
  Among them, we focus on the so-called $D_n$ theory which is associated with a sphere 
  with a regular puncture of degree $2$ and an irregular puncture of degree $n+2$.
  
  In section \ref{subsec:Dn} we will see that this theory is obtained as a deformation 
  of the maximally superconformal fixed point of $\CN=2$, $SU(n-1)$ gauge theory 
  with two fundamental hypermultiplets \cite{ARSW, EHIY}.
  Since $D_n$ theories have an $SU(2)$ flavor symmetry corresponding to the regular puncture,
  we can gauge it to obtain $SU(2)$ gauge theories coupled to them.
  In section \ref{subsec:Amn}, we will see that for one $SU(2)$ gauge group, at most two $D_n$ theories can be coupled
  and the corresponding Riemann surface is a sphere with at most two irregular punctures.
  We will call this as $\hat{A}_{m,n}$ theory following \cite{CV}.
  Finally, we will discuss more generic situations, namely $SU(2)$ wild quiver gauge theory
  associated with a Riemann surface with various regular and irregular punctures in section \ref{subsec:generalization}.
  The analysis in this section is purely field-theoretical.
  The geometric viewpoint from string and M-theory will be analyzed in the next section.
  
\subsection{$D_{n}$ theories}
\label{subsec:Dn}
  Let us first see that the Seiberg-Witten curve of the $D_{n}$ theory is realized as a double cover 
  of a sphere with an irregular puncture of degree $n+2$ and a regular one of degree $2$.
  Let us start with the Seiberg-Witten curve of $SU(n-1)$ gauge theory with two fundamental hypermultiplets
  which has $U(2)$ flavor symmetry \cite{SW1,SW2, APS, HO}
    \bea
    y^2
     =     (\hat{x}^{n-1} + \hat{u}_2 \hat{x}^{n-3} + \ldots + \hat{u}_{n-1})^2
         - \Lambda^{2n - 4} \prod_{i=1,2} (\hat{x} + m_i),
    \eea
  where $m_i$ are the mass parameters of the fundamentals and $\hat{u}_i$ are the Coulomb moduli.
  We follow the procedure used in \cite{GST} to obtain the maximally conformal point.
  Let us define $u_1 = - \frac{n-1}{2} (m_1 + m_2)$ and $C_2 = - \frac{1}{4}(m_1 - m_2)^2$
  which are related with the mass parameters associated with the $U(1)$ and the $SU(2)$ flavor symmetries respectively.
  We first shift $\hat{x} = x + \frac{u_1}{n-1}$ to obtain
    \bea
    y^2
     =     (x^{n-1} + u_1 x^{n-2} + u_2 x^{n-3} + \ldots + u_{n-1})^2 - \Lambda^{2n - 4} (x^2 + C_2),
           \label{curve2}
    \eea
  where $u_i$ with $i=2, \ldots, n-1$ are defined to include 
  the shifts due to the change of the coordinate $x$.
  Then, this curve can be written as
    \bea
    y^2
     =     (x^{n-1} + \ldots + \tilde{u}_{n-2} x + u_{n-1})
           (x^{n-1} + \ldots + (\tilde{u}_{n-2} - 2 \Lambda^{n-2})x + u_{n-1})
         - \Lambda^{2n-4} C_2,
    \eea
  where $\tilde{u}_{n-2} = u_{n-2} + \Lambda^{n-2}$.
  We can easily see that when the moduli $u_i$, $\tilde{u}_{n-2}$ and $C_2$ are small compared with $\Lambda$
  the curve around $x=0$ degenerates to 
    \bea
    y^2 
     \sim
         - 2 \Lambda^{n-2} x^{n},
    \eea
  which indicates the maximally conformal point.
  In this limit, the Seiberg-Witten differential can be written as
    \bea
    \lambda
     =     2 \hat{x} \frac{P^2}{M} d \left( \frac{y}{P} \right)
     \sim
           \frac{1}{\Lambda^{n-2}} \frac{y dx}{x},
    \eea
  where $P = \hat{x}^{n-1} + \hat{u}_2 \hat{x}^{n-3} + \ldots + \hat{u}_{n-1}$ and $M = P^2 - y^2$.
  
  To see a small deformation from this point, we define
  $\tilde{y}^2 = \frac{y^2}{\Lambda^{2n-4} x^2}$.
  The Seiberg-Witten curve of the deformed theory is 
    \bea
    \tilde{y}^2
     =     x^{n-2} + c_1 x^{n-3} + \ldots + c_{n-2} + \frac{c_{n-1}}{x} + \frac{C_2}{x^2},
           \label{Dn+1curve1}
    \eea
  with Seiberg-Witten differential $\lambda = \tilde{y} dx$.
  The parameters $c_i$ descend from the moduli parameters $u_i$ of the original curve.
  Note that we have rescaled $x$ and $u_i$ to get rid of the dynamical scale.
  It is obvious that the quadratic differential $\lambda^2 = \tilde{y}^2 (dx)^2$ has
  a pole of degree $2$ at $x=0$ and a pole of degree $n+2$ at $x = \infty$.
  Thus, we can see that similarly to Gaiotto's construction, 
  the Seiberg-Witten curve (\ref{Dn+1curve1}) is a double cover of the sphere with one regular puncture 
  and one degree $n+2$ puncture, which is the curve of the $D_{n}$ theory.
  The topology of the curve is as follows: if we define $\tilde{y}^2 = x^{-2} P_{n}(z)$, 
  then the branch points are at the roots of $P_{n}$ for even $n$ and at the roots of $P_{n}$ and $\infty$ for odd $n$.
  Therefore, the genus of the Seiberg-Witten curve is $\frac{n}{2}-1$ for $n$ even and $\frac{n-1}{2}$ for $n$ odd.
  
  The scaling dimensions of the deformation parameters can be easily determined 
  by demanding $\Delta(\lambda) = 1$.
  It follows that 
    \bea
    \Delta(C_2)
     =     2, ~~~
    \Delta(c_i)
     =     \frac{2i}{n}, ~~i=1, \ldots, n-1
           \label{dimensionsDn}
    \eea
  Let us consider the meaning of these dimensions.
  In general, from a non-trivial superconformal fixed point on the Coulomb branch in an $\CN=2$ gauge theory, 
  one can consider a deformation by adding to the Lagrangian
  $\delta \CL = \int d^2 \theta_1 d^2 \theta_2 m V$,
  where $\theta_{1,2}$ are the superspace variables of $\CN=2$ supersymmetry
  and $V$ is the $\CN=2$ vector superfield whose lowest component is the chiral primary field $v$.
  Since we consider a non-trivial CFT, $v$ is a relevant operator if $1 < \Delta(v) \leq 2$,
  and equivalently $m$ is a relevant parameter if $0 \leq \Delta(m) < 1$.
  Then, returning to the $D_n$ theory, one can find a remarkable relation between the dimensions: 
  $\Delta(c_i) + \Delta(c_{n-i}) = 2$ and when $n$ is even $\Delta(c_{n/2}) = 1$.
  This implies that $c_i$ for $i= [\frac{n}{2}] + 1, \ldots, n-1$, where $\Delta(c_i) > 1$,
  are the vevs of the relevant operators while
   for $i= 1, \ldots, [\frac{n}{2}]$, where $\Delta(c_i) \leq 1$, the coefficients $c_i$ are their corresponding
  couplings and a dimension $1$ parameter when $n$ is even\footnote{$[s]$ is the integer part of $s$.}.
  The number of relevant operators is therefore $[\frac{n-1}{2}]$.
  In order to make the difference explicit, we rename the vevs as $v_a$
  where $a = 1, \ldots, [\frac{n-1}{2}]$ in the following.
  
  The parameter $C_2$, whose dimension is $2$, comes from the Casimir of the mass parameter 
  associated with the $SU(2)$ flavor symmetry.
  Since the superconformal fixed point keeps this $SU(2)$ flavor symmetry,
  $C_2$ is still the Casimir mass parameter.
  Indeed, for a mass parameter associated with a non-Abelian flavor symmetry,
  the scaling dimension does not acquire an anomaly \cite{ARSW}.
  Note that when $n=3$ the original flavor symmetry is enhanced to $SO(4) \sim SU(2) \times SU(2)$.
  However, the fixed point preserves only one of the $SU(2)$'s, because of $u_1 \sim \Lambda$ at that point.
  (Note that we have defined $\tilde{u}_1 = u_1 + \Lambda$ for $n=3$.)
  
  Note also that these fixed points are in the same universality class of the ones
  which can be obtained from $\CN=2$, $SO(2n)$ pure Yang-Mills theory 
  as maximally conformal points called as $MD_n$ in \cite{EHIY}.
  In fact, the dimensions of the parameters (\ref{dimensionsDn}) are the same as those in the table 3 in \cite{EHIY}.
  The name $D_n$ seems to be more suitable from this viewpoint.
  However, from the $SO(2n)$ gauge theory viewpoint, it is not trivial to see how the $SU(2)$ flavor symmetry arises,
  though this was trivially seen in the analysis in this subsection.
  
  For future reference, let us consider the curve (\ref{Dn+1curve1}) 
  in the $w = \frac{1}{x}$ coordinate:
    \bea
    y^2
     =     \frac{1}{w^{n+2}} + \frac{c_1}{w^{n+1}} + \ldots + \frac{c_{[\frac{n}{2}]}}{w^{[\frac{n+3}{2}]+1}} 
         + \frac{v_{[\frac{n-1}{2}]}}{w^{[\frac{n+3}{2}]}} + \ldots + \frac{v_1}{w^3}
         + \frac{C_2}{w^2},
           \label{Dn+1curve2}
    \eea
  with the Seiberg-Witten differential $\lambda = y dw$.
  We call this as $D_{n}$ curve.
  Note that we have assumed from the beginning that $n \geq 3$.
  However, even for $n=1,2$ the curve exists, although this does not describe a nontrivial theory.
  We will see these explicitly below.
  
  The flavor central charges of these theories have been computed in \cite{AT, GST} as
   \bea
   k
    =     \frac{4(n-1)}{n}.
          \label{centralDn+1}
   \eea
  We will use this later.

\subsection{$\hat{A}_{m,n}$ theories}
\label{subsec:Amn}
  Now, we consider $\hat{A}_{m,n}$ theories.
  As already stated in \cite{CV}, this is an $SU(2)$ gauge theory 
  coupled to two SCFTs $D_m$ and $D_n$.
  Let us first consider the small $m$ and $n$ cases.

\subsubsection*{$\hat{A}_{1,1}$ theory}
  The simplest one is $\hat{A}_{1,1}$ which is $SU(2)$ pure super Yang-Mills theory.
  The curve is $x^2 = \phi_2$ with \cite{GMN}
    \bea
    \phi_2
     =     \frac{\Lambda^2}{z^3} + \frac{u}{z^2} + \frac{\Lambda}{z},
           \label{curveA11}
    \eea
  where $u$ is the Coulomb moduli parameter.
  The Seiberg-Witten differential is denoted as $\lambda = x dz$.
  The quadratic differential $\phi_2 (dz)^2 = \lambda^2$ has poles of degree $3$ at $z=0$ and $\infty$.
  The Seiberg-Witten curve is a double cover of the sphere with two irregular punctures of degree $3$.
  
  We can obtain the $D_1$ curve by taking the weak coupling limit $\Lambda\to 0$ of the above.
  By redefining $z = \Lambda^2 w$, we obtain 
  $\tilde{x}^2 = \tilde{\phi}_2 = \Lambda^4 \phi_2 \rightarrow \frac{1}{w^3} + \frac{u}{w^2}$,
  where the differential is once again $\lambda = \tilde{x} dw$.
  This is the $D_1$ curve (\ref{Dn+1curve2}).
  Conversely, we can obtain the $SU(2)$ pure Yang-Mills theory by gauging the diagonal $SU(2)$ flavor symmetries
  of two $D_1$ theories.
  However, these theories are empty and do not contribute to the beta function of the coupled $SU(2)$ gauge theory.
  
\subsubsection*{$\hat{A}_{1,2}$ theory}
  This is $SU(2)$ gauge theory with one fundamental hypermultiplet.
  The curve is given by \cite{GMN}
    \bea
    \phi_2
     =     \frac{\Lambda^2}{z^4} + \frac{m \Lambda}{z^3} + \frac{u}{z^2} + \frac{\Lambda^2}{z}.
           \label{curveA21}
    \eea
  The quadratic differential has poles of degree $4$ at $z=0$ and of degree $3$ at $z = \infty$.
  Thus, this theory is associated with the sphere with two punctures of degree $4$ and $3$.
  
  Similarly to the previous case, we can obtain the $D_2$ curve from the above one by redefining $z = \Lambda w$ 
  and taking $\Lambda \rightarrow 0$. In other words, by looking at the region near $z = 0$ we get
  $\tilde{\phi}_2 = \Lambda^2 \phi_2 \rightarrow \frac{1}{w^4} + \frac{m}{w^3} + \frac{u}{w^2}$.
  Indeed the $D_2$ theory is that of four free half-hypermultiplets and 
  contributes to the one-loop beta function coefficient by $-1$ once the $SU(2)$ symmetry is gauged.
  
\subsubsection*{$\hat{A}_{1,3}$ theory}
  This is the first example which includes a nontrivial SCFT.
  By generalizing the curve (\ref{curveA21}), we obtain the Seiberg-Witten curve $x^2 = \phi_2$ where
    \bea
    \phi_2
     =     \frac{\Lambda^2}{z^5} + \frac{\Lambda^{\frac{4}{3}} c_1}{z^4}
         + \frac{\Lambda^{\frac{2}{3}} v_1}{z^3} + \frac{u}{z^2} + \frac{\Lambda^2}{z}.
           \label{Ahat31curve}
    \eea
  This theory is the $SU(2)$ gauge theory coupled to the $D_3$ theory.
  Indeed, by redefining $z = \Lambda^{2/3} w$ and taking the weak coupling limit $\Lambda \rightarrow 0$, we obtain 
    \bea
    \tilde{\phi}_2
     =     \Lambda^{4/3} \phi_2
     \rightarrow
           \frac{1}{w^5} + \frac{c_1}{w^4} + \frac{v_1}{w^3} + \frac{u}{w^2},
    \eea
  which is the $D_3$ curve (\ref{Dn+1curve2}).
  As can be expected, 
  the Coulomb moduli parameter $u$ corresponds to the mass parameter $C_2$ associated with the $SU(2)$ flavor symmetry 
  after decoupling.
  
  The Seiberg-Witten curve of the $\hat{A}_{1,3}$ theory (\ref{Ahat31curve}) is a double cover 
  of the sphere with two punctures of degree $5$ and $3$.
  The branch points are at $z=0$, $\infty$ and at the four roots of the polynomial $P_4(z)$, 
  where we defined $\phi_2 = z^{-5} P_4(z)$.
  Therefore the genus of the Seiberg-Witten curve is two.
  This matches with the fact that this theory can be seen as a deformation of the superconformal point 
  of a parent $SU(2) \times SU(2)$ quiver gauge theory with one bifundamental hypermultiplet,
  whose Seiberg-Witten curve is indeed genus $2$.
  
  One can also see that the derivatives of the Seiberg-Witten differential $\lambda$
  with respect to $u$ and $\Lambda^{\frac{2}{3}} v_1$ give 
  a basis of holomorphic differentials on the curve.
  Note that the derivative with respect to $c_1$ does not give an independent holomorphic differential
  since $c_1$ corresponds to a mass parameter not associated to the moduli 
  of the original $SU(2) \times SU(2)$ theory.
  
  Since the $D_3$ sector contributes to the one-loop beta function coefficient of the $SU(2)$ gauge theory
  by $-\frac{k}{2} = -\frac{4}{3}$, where $k$ is given by (\ref{centralDn+1}), 
  this coupled theory has $b_0 =  \frac{8}{3}$.
  This fractional number reflects the fractional power of $\Lambda$ in (\ref{Ahat31curve}).

\subsubsection*{$\hat{A}_{m,n}$ theory}
  It is straightforward to generalize the above construction to the $\hat{A}_{1,n}$,
  that is to the $SU(2)$ gauge theory coupled to the $D_{n}$ SCFT.
  The curve is
    \bea
    \phi_2
     =     \frac{\Lambda^2}{z^{n+2}} + \frac{\Lambda^{\frac{2n-2}{n}} c_1}{z^{n+1}} 
         + \frac{\Lambda^{\frac{2n -4}{n}} c_2}{z^{n}} + \ldots + \frac{\Lambda^{\frac{2}{n}} v_{2}}{z^4} 
         + \frac{\Lambda^{\frac{2}{n}} v_{1}}{z^3}
         + \frac{u}{z^2} + \frac{\Lambda^2}{z}
    \eea
  which is associated to a sphere with punctures of degree $n+2$ and $3$.
  Indeed, by taking the weak coupling limit with $z = \Lambda^{\frac{2}{n}} w$, 
  we reproduce the $D_n$ curve (\ref{Dn+1curve2}).
  Let us check the genus of this curve.
  We define $\phi_2 = z^{-(n+2)} P_{n+1} (z)$ where $P_{n+1}$ is a polynomial of degree $n+1$.
  When $n$ is even, the branch points of the curve are at the roots of $P_{n+1}$ and at $z=\infty$,
  thus the genus is $\frac{n}{2}$.
  When $n$ is odd, the branch points are at the roots and $z=0, \infty$, leading to genus $\frac{n+1}{2}$.
  This is greater than the genus of the $D_n$ curve by one, as can be expected.
  
  Again, the one-loop beta function coefficient of this $SU(2)$ gauge theory is
    \bea
    b_0
     =   \frac{2 n + 2}{n}.
    \eea
  We can expect that this theory can be obtained as a fixed point of 
  $SU(n-1) \times SU(2)$ quiver gauge theory
  with a bifundamental field in ($\textrm{\boldmath $n-1$}, \textrm{\boldmath $2$}$) representation.
  
  Now, we consider the most general case $\hat{A}_{m,n}$
  which is the $SU(2)$ gauging of the $D_{n}$ and $D_{m}$ SCFTs.
  The corresponding curve is
    \bea
    \phi_2
    &=&    \frac{\Lambda^2}{z^{n+2}} + \frac{\Lambda^{\frac{2n-2}{n}} c_1}{z^{n+1}} 
         + \frac{\Lambda^{\frac{2n -4}{n}} c_2}{z^{n}} + \ldots + \frac{\Lambda^{\frac{2}{n}} v_{1}}{z^3}
           \nonumber \\
    & &    ~~~~~~
         + \frac{u}{z^2} + \frac{\Lambda^{\frac{2}{m}} \tilde{v}_{1}}{z} + \ldots
         + \Lambda^{\frac{2m-2}{m}} \tilde{c}_1 z^{m-3} + \Lambda^2 z^{m-2},
    \eea
  where $c_i$ and $v_a$ ($\tilde{c}_j$ and $\tilde{v}_b$) are the relevant parameters and 
  the vevs of the relevant operators of the $D_n$ ($D_m$) theory respectively.
  The one-loop beta function coefficient is $b_0 = 2 (\frac{1}{n} + \frac{1}{m})$.
  
  At this stage, let us count the number of moduli of this theory.
  Obviouslus there is a single Coulomb modulus while the other parameters
  are associated to the vevs of the relevant operators $v_a$.
  As found in the previous subsection, the number of them for the $D_n$ theory is $[\frac{n-1}{2}]$.
  Therefore, the total number of the moduli is 
    \bea
    (\# {\rm~of ~moduli})
     =     1 + \left[ \frac{m-1}{2} \right] + \left[ \frac{n-1}{2} \right].
           \label{numbermoduliA}
    \eea
   
  When $n=2$ or $m =2$, the coupled sector is just a fundamental hypermultiplet.
  Hence the $\hat{A}_{n,2}$ theory is an $SU(2)$ gauge theory coupled to the $D_n$ SCFT
  and one fundamental hypermultiplet.
  The parameter $\tilde{c}_1$ in this case is nothing but the mass parameter of the hypermultiplet.
  Finally, when $n=m=2$, this reduces to the $SU(2)$ gauge theory with two fundamental hypermultiplets
  whose Seiberg-Witten curve is described by \cite{GMN}
    \bea
    \phi_2
     =     \frac{\Lambda^2}{z^4} + \frac{m \Lambda}{z^3} + \frac{u}{z^2} + \frac{\tilde{m} \Lambda}{z} + \Lambda^2.
           \label{curveA22}
    \eea
  Of course, the number of moduli (\ref{numbermoduliA}) is one for the $m=n=2$ case.
  
  In summary, we constructed the Seiberg-Witten curve of the $SU(2)$ gauge theory coupled to two SCFTs
  by using as building blocks the $D_{n}$ theories, the theory being associated to the sphere with two irregular punctures.
  It is also possible to consider gauge theories associated 
  to generic Riemann surfaces with many irregular and regular punctures.
  This turns out to be an $SU(2)$ wild quiver gauge theory.
  We will see this below.

\subsection{Generalization to wild quivers}
\label{subsec:generalization}
  In this subsection we generalize the above analysis to Riemann surface with 
  $\ell$ regular punctures and $k$ irregular punctures with degree $n_\alpha + 2$ ($\alpha=1, \ldots, k$)
  which we denote as $\CC_{g,\ell, \{ n_\alpha \}}$.
  By definition, $n_\alpha \geq 1$. 
  
  The basic building blocks are $\CC_{0,3}$ and $\CC_{0,1,\{ n \}}$
  where $\CC_{0,3} = \CC_{0,3,\{ \emptyset \}}$ and $\emptyset$ denotes the absence of irregular punctures.
  The theory corresponding to $\CC_{0,3}$ is that of four free hypermultiplets 
  which was called $T_2$ theory in \cite{Gaiotto}.
  On the other hand, as analyzed in the previous section, 
  the latter is the new ingredient inducing the $D_{n}$ theory.
  We can construct a large class of $SU(2)$ wild quiver gauge theories 
  with (bi and tri)fundamental hypermultiplets and coupled to SCFTs,
  by gauging $SU(2)$ flavor symmetries of the $T_2$ and $D_{n}$ theories.
  This gauging process corresponds to connect regular punctures of 
  $\CC_{0,3}$'s and $\CC_{0,1, \{ n \}}$'s by thin tubes. 
  In this way one can get any Riemann surface $\CC_{g, \ell, \{ n_\alpha \}}$.
  
  Let us consider these quiver gauge theories more explicitly starting with the $g=0$ case.
  According to our general construction, 
  the (effective) beta function coefficients could be vanishing or positive.
  Actually, connecting two $\CC_{0,3}$'s leads to an $SU(2)$ gauge group 
  whose one-loop beta function coefficient is zero.
  Connecting $\CC_{0,3}$ and $\CC_{0, 1, \{ n \}}$ gives rise to an asymptotically free gauge group.
  Therefore, the number of the asymptotically free $SU(2)$ gauge groups is $k$,
  the number of the irregular singularities\footnote{This counting changes in the  $\hat{A}_{m,n}$ case  which is 
  obtained by connecting two $\CC_{0,1,\{ m \}}$ and $\CC_{0,1,\{n\}}$.}.
  Then, there are $\ell + k -3$ $SU(2)$ gauge groups which have vanishing beta function coefficients.
  The (bi and tri)fundamental hypermultiplets couple with these gauge groups
  preserving a total flavor symmetry $SU(2)^\ell$.
  
  Correspondingly, $\ell+k-3$ complex structure moduli of the Riemann surface $\CC_{0, \ell, \{ n_\alpha \} }$ 
  are identified with the gauge coupling constants of the gauge groups with vanishing beta function coefficients 
  $q_i = e^{2 \pi i \tau_i}$.
  The Seiberg-Witten curve is the double cover of this sphere: $x^2 = \phi_2(z)$.
  The quadratic differential locally has the following structures:
  at the regular punctures $z = z_f$ ($f=1, \ldots, \ell$),
    \bea
    \phi_2
     \sim
           \frac{m_f^2}{(z - z_f)^2} + \ldots,
           \label{localSW1}
    \eea
  The residues of $\lambda$ at $z = z_f$ are the mass parameters associated with the $SU(2)^\ell$ flavor symmetry.
  Near the irregular punctures, $z = z_\alpha$ ($\alpha = 1, \ldots, k$)
    \bea
    \phi_2
     \sim  \frac{\Lambda_\alpha^2}{(z - z_\alpha)^{n_\alpha +2}}
         + \frac{\Lambda_\alpha^{\frac{2n_\alpha-2}{n_\alpha}} c_{1}^\alpha}{(z - z_\alpha)^{n_\alpha+1}} 
         + \frac{\Lambda^{\frac{2n_\alpha -4}{n_\alpha}}_\alpha c_{2}^\alpha}{(z - z_\alpha)^{n_\alpha}} + \ldots
         + \frac{\Lambda^{\frac{2}{n_\alpha}}_\alpha v_{1}^\alpha}{(z - z_\alpha)^3}
         + \frac{u_\alpha}{(z - z_\alpha)^2}
         + \ldots
           \label{localSW2}
    \eea
  where $u_\alpha$ and $\Lambda_\alpha$ are, respectively, the Coulomb moduli and the dynamical scale 
  of the gauge group which couples to the $D_{n_{\alpha}}$ theory, and
  $c_{i}^{\alpha}$ ($i= 1, \ldots, [\frac{n_\alpha}{2}]$) and $v_{a}^\alpha$ ($a=1, \ldots, [\frac{n_\alpha-1}{2}]$) 
  are the parameters labeling the deformations from the fixed point.
  The other Coulomb moduli parameters are encoded in the less singular terms in $\phi_2$.
  
  The case with $\ell=0$ and $k=2$ is exceptional in the sense 
  that the above counting of the number of the gauge groups is invalid.
  This corresponds to the $\hat{A}_{m,n}$ theory analyzed in the previous section.
  The case with $\ell=2$ and $k=1$ was analyzed in \cite{CV} where it was called $\hat{D}_n$ theory.
  These are the only two cases having one asymptotically free $SU(2)$ gauge group.
  
  For $g>0$, the construction is similar to the above one.
  The number of asymptotically free gauge groups is still equal to $k$, and
  the number of gauge groups with vanishing beta function coefficient is $3g - 3 + \ell + k$, 
  which agrees with the number of complex structure moduli of the Riemann surface $\CC_{g, \ell, \{ n_\alpha \}}$.
  There are in total $3 g - 3 + \ell + 2k$ Coulomb moduli of the $SU(2)$ gauge groups.
  Let us now include in the counting the vevs of the relevant operators $v_a^\alpha$.
  Therefore, the total number of moduli is
    \bea
    (\# ~{\rm of} ~{\rm the ~moduli})
     =     3g - 3 + \ell + 2k + \sum_{\alpha = 1}^k \left[ \frac{n_\alpha - 1}{2} \right].
           \label{numbermoduli}
    \eea
  The Seiberg-Witten curve is given by a double cover of $\CC_{g, \ell, \{ n_\alpha \}}$
  and its local behavior at the singularities is that of (\ref{localSW1}) or (\ref{localSW2}).

\section{Geometric interpretation}
\label{sec:geom}
  So far, we considered four-dimensional $SU(2)$ wild quiver gauge theories 
  from a purely field theoretical point of view.
  Linear and elliptic quivers are induced 
  as world-volume theories of an appropriate intersecting D4-NS5 brane system.
  Its M-theory lift leads to two M5-branes wrapping the corresponding Riemann surface which is the base
  of the Seiberg-Witten double cover \cite{Witten, Gaiotto}.
  As found in \cite{Gaiotto}, even if type IIA brane configuration does not exist,
  a large class of superconformal quiver gauge theories can be obtained 
  by wrapping M5-branes on $\CC_{g, \ell, \{ \emptyset \}}$, that is the one with only regular singularities.
  More precisely, the theory is superconformal only at the origin of the moduli space
  and with vanishing masses.
  The analysis in the previous section suggests that the $SU(2)$ wild quiver gauge theory with $D_n$ sectors
  can also be obtained from two M5-branes compactified on $\CC_{g,\ell, \{ n_\alpha \}}$.
  
  The low energy dynamics of two M5-branes, after decoupling the center of mass mode, is governed 
  by the $\CN=(2,0)$ $A_1$ theory in six dimensions.
  Thus, the gauge theory constructed in the previous section is given
  by compactifying the $A_1$ $(2,0)$ theory on $\mathbb{R}^{1,3} \times \CC_{g,\ell,\{ n_\alpha \}}$.
  
  In this section, we develop the geometrical interpretation of $SU(2)$ wild quiver gauge theories.
  In order to do that, it is crucial to find the related integrable system. 
  As discussed in \cite{DW} and more recently in \cite{GMN, NX1}, 
  for a large class of $\CN=2$ superconformal quiver gauge theories obtained from the $(2,0)$ theory on
  $\mathbb{R}^{1,3} \times \CC_{g,\ell, \{ \emptyset \}}$, 
  the Seiberg-Witten fibration was identified with the Hitchin integrable system (or Hitchin fibration)
  associated to $\CC_{g,\ell, \{ \emptyset \}}$ \cite{Hitchin, Hitchin1, Simpson, Markman, DonagiMarkman}.
  The singularity in the $SU(2)$ superconformal case is the mildest one:
  we allow at most double poles of the quadratic differential $\phi_2$.
  In terms of the Hitchin moduli space, this corresponds to tame ramifications 
  where the gauge and Higgs fields have simple poles.
  However, we can allow a higher order singularity of the Higgs field,
  which is called wild ramification \cite{BB}.
  Therefore, it is natural to expect that our theory associated with $\CC_{g,\ell,\{ n_\alpha \}}$
  where we allow higher order singularities is related to the Hitchin moduli space with wild ramifications.
  The case with $g=0$, $\ell=0$ and $n_\alpha\leq 2$ has been already discussed in \cite{NX2}.
  
  Let us explain why such a connection with the Hitchin moduli space appears, 
  starting from the $(2,0)$ theory in six dimensions.
  As argued above, the $(2,0)$ theory compactified on a Riemann surface induces 
  a four-dimensional $\CN=2$ gauge theory.
  Furthermore, let us consider its compactification on $\mathbb{R}^{1,1} \times S^1 \times S^1$.
  By compactifying on $S^1$, we get a three-dimensional gauge theory 
  whose Coulomb moduli space $\CM$ is an hyper-K\"ahler manifold.
  This, in a particular complex structure, is the Seiberg-Witten fibration of the four-dimensional theory \cite{SW3d}.
  By further compactifying on $S^1$, we are led to
  a two-dimensional $\CN= (4,4)$ sigma model with target space $\CM$.
  Let us go back to the $(2,0)$ theory and reverse the order of the compactifications, 
  namely we first compactify the $(2,0)$ theory on $S^1 \times S^1$, 
  which leads to $\CN=4$ super Yang-Mills theory in four dimensions.
  Then, by compactifying on the Riemann surface with a suitable twist, 
  we get a sigma model whose target space is the Hitchin moduli space 
  $\CM_{{\rm H}}$ \cite{HMS, BJSV, KW, GW, WildWitten}.
  Therefore, comparing the two perspectives of the compactification of the $(2,0)$ theory suggests 
  a relation between the low energy physics of four-dimensional $\CN=2$ gauge theory and the Hitchin moduli space.
  Note that a similar argument for the case of $\mathbb{R}^{1,2} \times S^1$ also leads 
  to the same conclusion \cite{CK, Kapustin, GMN}.
  
  We will see below that wild quiver gauge theories are indeed related to Hitchin systems with wild ramifications,
  and that this gives a further geometric understanding of the gauge theory.
  We first give a review of the Hitchin system with wild ramifications in section \ref{subsec:Hitchin}.
  Then, we discuss the correspondence with the $D_n$ theories in section \ref{subsec:HitchinDn} and finally,
  in section \ref{subsec:generalHitchin}, we describe the wild quiver gauge theory
  in terms of the Hitchin systems with wild ramifications.

\subsection{Hitchin system with wild ramifications}
\label{subsec:Hitchin}
  In this subsection, we review the Hitchin system with wild ramifications.
  While the gauge theory considered above corresponds to the Hitchin moduli space with the gauge group $SU(2)$,
  here we discuss the case of a generic gauge group.
    
  First of all, we consider the case without ramification.
  Let $G$ be a Lie group whose algebra is denoted by $\mathfrak{g}$.
  Let $E$ be  a $G$-bundle on a Riemann surface $\CC_g \equiv \CC_{g, 0, \{ \emptyset \}}$ with a connection $A$
  and $\phi$ be a one-form valued in ${\rm ad}(E)$.
  The space parametrized by $(A, \phi)$ has an hyper-K\"ahler structure,
  with three complex structures conventionally written as $I$, $J$ and $K$
  satisfying $IJ = K$.
  In particular, in the complex structure $I$, $A_{\bar{z}}$ and $\phi_z$ are holomorphic, 
  $A_{\bar{z}}$ and $\phi_z$ being the $(0,1)$ and $(1,0)$ components of $A$ and $\phi$ respectively.
  In the complex structure $J$ instead, $A_z + i \phi_z$ and $A_{\bar{z}} + i \phi_{\bar{z}}$ are holomorphic.
  In other words, the $G_{\mathbb{C}}$ valued connection $\CA = A + i \phi$ is holomorphic.
  This implies that $J$ does not depend on the complex structure of $\CC_g$.
  \footnote{We are following here the notation in \cite{KW}.}
  Correspondingly, there are three symplectic structures $\omega_I$, $\omega_J$ and $\omega_K$.
  Let us define $\Omega_I = \omega_J + i \omega_K$, and $\Omega_J$ and $\Omega_K$ as its cyclic permutations.
  In this notation, $\Omega_I$ and $\Omega_J$ can be written as
    \bea
    \Omega_I
     \sim
           \int_{\CC_g} d^2 z \Tr \delta \phi_z \wedge \delta A_{\bar{z}},~~~~
    \Omega_J
     \sim
           \int_{\CC_g} d^2 z \Tr \delta \CA \wedge \delta \CA,
           \label{OmegaI}
    \eea
  where $\delta$ denotes the exterior derivative on the space of $(A, \phi)$.
  These are holomorphic (2,0) forms in the complex structures $I$ and $J$ respectively.
  
  The Hitchin equations are 
    \bea
    F - \phi \wedge \phi
    &=&    0,
           \nonumber \\
    D \phi
     =     D \star \phi
    &=&    0
           \label{Hitchineq}
    \eea
  where $F$ is the curvature of the connection and $\star$ is the Hodge star.
  The Hitchin moduli space $\CM_{{\rm H}}^{reg}$ is the set of regular solutions to (\ref{Hitchineq}) 
  divided by $G$ gauge transformations.
  
  Let us consider $\CM_{{\rm H}}^{reg}$ in the complex structure $I$.
  The Hitchin equations in the second line of (\ref{Hitchineq}) are equivalent 
  to $D_{\bar{z}} \phi_z = 0$ and its complex conjugate.
  This is holomorphic in $I$ since this only depends on $A_{\bar{z}}$ and $\phi_z$.
  (This equation is equivalent to the vanishing of the moment map associated with $\Omega_I$.)
  On the other hand, the first equation is a ``real" equation
  (which is equivalent to the vanishing of the real moment map with $\omega_I$).
  It turns out to be convenient to treat these equations separately.
  Let us define $\varphi = \phi_z dz$ and call it as the Higgs field.
  The equation $D_{\bar{z}} \phi_z = 0$ means that $\varphi$ is a holomorphic section of $K_{\CC_g} \otimes {\rm ad}(E)$
  where $K_{\CC_g}$ is the canonical line bundle on $\CC_g$.
  Thus, the solutions to the holomorphic equation are described by a pair $(E, \varphi)$ 
  where $E$ is a holomorphic $G$-bundle determined by $A_{\bar{z}}$.
  
  Then, we impose the real equation $F - \phi \wedge \phi = 0$.
  It can be shown that imposing this equation and quotienting by $G$ is equivalent 
  to quotienting the pair $(E, \varphi)$
  by complexified $G_{\mathbb{C}}$ gauge transformations, modulo stability.
  Summarizing, the Hitchin moduli space $\CM_{{\rm H}}$, in the complex structure $I$, 
  is the pair $(E, \varphi)$ divided by $G_{\mathbb{C}}$.
    
  A similar analysis can be applied to the system in the complex structure $J$.
  The result is that $\CM_{{\rm H}}^{reg}$ is 
  the space of $G_{\mathbb{C}}$ flat connections $\CA$ divided by $G_{\mathbb{C}}$ gauge transformations,
  again modulo stability.
  Indeed, the vanishing of the moment map associated with $\Omega_J$ is equivalent to the flatness condition.
  The complex dimension of the moduli space is $\dim_{\mathbb{C}} \CM_{{\rm H}}^{reg} = 2(g-1) \dim (G)$.
  Since $\CA$ is a flat connection, the moduli space can be specified by 
  the monodromies around the independent $A$ and $B$ cycles of $\CC_g$, $A_\alpha$ and $B_\alpha$ ($\alpha= 1, \ldots, g$).
  These are $G_{\mathbb{C}}$ valued and satisfy the condition:
    \bea
    1 
     =     A_1 B_1 A_1^{-1} B_1^{-1} \cdot \cdot \cdot A_g B_g A_g^{-1} B_g^{-1}.
           \label{monodromyflatconnection}
    \eea
  By dividing the $G_{\mathbb{C}}$ gauge transformations, one can obtain the dimension above.

  Let us now go back to the complex structure $I$ and describe the so-called Hitchin fibration
  as a completely integrable system \cite{Hitchin}.
  For simplicity and for the purpose of this paper, we choose $G = SU(2)$.
  Then, we consider the space of gauge invariant polynomials of $\varphi$. 
  In the $SU(2)$ case, this is generated by $\Tr \varphi^2$, a holomorphic quadratic differential 
  parametrizing $\CQ = H^0(\CC_g, K_{\CC}^2)$.
  The Hitchin fibration is specified by a map $\CM_{{\rm H}}^{reg} \rightarrow \CQ$.
  The complex dimension of the base space $\CQ$ is simply given by $3(g-1)$
  which is one half of the complex dimension of $\CM_{{\rm H}}^{reg}$.
  
  The commuting Hamiltonians $H_p$ ($p = 1, \ldots, 3(g-1)$) can be constructed from $\Tr \varphi^2$ as
    \bea
    H_p
     =     \int_{\CC_g} \alpha_p \wedge \Tr \varphi^2,
           \label{Hamiltonians}
    \eea
  where $\alpha_p$ are basis of Beltrami differentials.
  We can easily show that these Hamiltonians commute with each other
  with respect to the holomorphic symplectic form $\Omega_I$ (\ref{OmegaI}),
  because (\ref{Hamiltonians}) depends only on $\phi_z$.
  The spectral curve of the integrable system is defined by 
    \bea
    x^2
     =     \Tr \varphi^2,
           \label{spectral}
    \eea
  where we omitted $(dz)^2$ and considered $\Tr \varphi^2$ as the coefficient of the quadratic differential.
  This is identified with the Seiberg-Witten curve of the corresponding $SU(2)$ quiver gauge theory.
  In particular, $\Tr \varphi^2$ is identified with $\phi_2$ in the Seiberg-Witten curve in section \ref{sec:gauge}.

\subsubsection*{Wild ramification}
  Let us consider now the singular solutions.
  We focus on the case with one singularity of degree $m$ at $z = 0$ 
  where $z$ is a local coordinate on the Riemann surface.
  The generalization to more singularities is straightforward, as we will see in subsequent subsections.
  
  Let $\mathbb{T}$ be the maximal torus of $G$.
  Let also $\mathfrak{t}$ and $\mathfrak{t}_{\mathbb{C}}$ be the Lie algebras of $\mathbb{T}$
  and its complexification respectively.
  We define the parameters $t_i \in \mathfrak{t}_{\mathbb{C}}$ and $\alpha \in \mathfrak{t}$.
  The singular solution which we focus on here is
    \bea
    A
    &=&    \alpha d \theta + \ldots,
           \nonumber \\
    \phi
    &=&    dz \left( \frac{t_m}{z^m} + \ldots + \frac{t_1}{z} \right) + \ldots + c.c.,
           \label{irregularsol}
    \eea
  where the ellipsis denotes the regular part.
  The moduli space $\CM_{{\rm H}}$ is given by a space parametrized by $(A, \phi)$ 
  divided by $G$ gauge transformations preserving the singular structure (\ref{irregularsol}).
  As in the case of the regular solutions, we consider the moduli space $\CM_{{\rm H}}$
  in the complex structures $I$ and $J$.
  
  In the complex structure $J$, the moduli space is a space of 
  a $G_{\mathbb{C}}$ valued flat connection $\CA$ with a singularity at $z=0$.
  $\CA$ can be transformed to the local form 
  $\CA = dz \left( \frac{2t_m}{z^m} + \ldots + \frac{2t_2}{z^2} - i \frac{\alpha - i 2\Im t_1}{z} \right)$.
  The moduli space is again parametrized by the monodromies.
  However, compared to the regular case, the inclusion of the singularity induces additional monodromy factors 
  to (\ref{monodromyflatconnection}) which are basically written in terms of the Stokes matrices.
  By counting the dimension of them, one obtains \cite{WildWitten}
    \bea
    \dim_{\mathbb{C}} \CM_{{\rm H}}
     =     2(g-1) \dim(G) + m (\dim(G) - r).
           \label{dimensionwitten}
    \eea
  where $r$ is the rank of $G$.
  Note that the contribution of the singularity corresponds to the second term in (\ref{dimensionwitten}).
  
  In the complex structure $I$, 
  we give a local trivialization of $E$ which reduces 
  $\bar{D} = d \bar{z} (\partial_{\bar{z}} + A_{\bar{z}})$
  to the $\bar{\partial}$ operator.
  The Higgs field $\varphi$ is given by the holomorphic part of (\ref{irregularsol})
    \bea
    \varphi
     =     dz \left( \frac{t_m}{z^m} + \ldots + \frac{t_1}{z} \right) + \ldots.
    \eea
  Therefore, the moduli space is described by the pair $(E, \varphi)$ where $E$ is a holomorphic $G$-bundle.
  This moduli space in this complex structure $I$ varies holomorphically with the parameters $t_1, \ldots, t_m$.
  
  Now, let us consider the Hitchin fibration $\CM_{{\rm H}} \rightarrow \CQ$
  where $\CQ$ is the space of a quadratic differential, focusing again to the $SU(2)$ case.
  The quadratic differential is locally
    \bea
    \Tr \varphi^2
     =     \frac{\Tr t_m^2}{z^{2m}} + \frac{2 \Tr t_m t_{m-1}}{z^{2m -1}} + \ldots
         + \frac{2 \Tr t_m t_1 + \ldots}{z^{m+1}} + \ldots.
           \label{quadratic}
    \eea
  Note that the terms less singular than $1/z^{m+1}$ depend on the regular terms of the Higgs field.
  This will be very important to make a connection to the gauge theory.
  The base space $\CQ$ is of complex dimension $3g-3 + m$.
  Indeed, this can be seen as follows: the parameters in $\CQ$ are the ones needed to specify
  the last dots in (\ref{quadratic}), since the more singular terms are fixed by $t_i$'s.
  This is one half of the complex dimension of $\CM_{{\rm H}}$ counted above.

\subsection{$D_n$ theory and Hitchin system}
\label{subsec:HitchinDn}
  We are ready to describe the singularity structure of the Higgs field
  which corresponds to the $D_n$ theory.
  Here we focus on the Hitchin fibration in the complex structure $I$.
  
  Before going into the $D_n$ theory, 
  let us briefly recall the case for $\CN=2$ superconformal $SU(2)$ gauge theories
  associated to the Riemann surface $\CC_{g,\ell}$.
  In this theory the relation with the Hitchin system is as follows:
  the Coulomb moduli space, parametrized by $u_i = \left< \tr \phi^2_i \right>$, is identified 
  with the base space $\CQ$ of the Hitchin fibration, as in \cite{DW}.
  Note that mass parameters do not correspond to the variables parametrizing $\CQ$.
  Indeed, the quadratic differential has at most double poles 
  which correspond to regular singularities of degree $1$ of the Higgs field.
  Thus, the parameter $t_1$ of the Higgs field is related to
  the mass parameter which is the residue of the Seiberg-Witten differential.
  As argued above, the Hitchin fibration depends holomorphically on this parameter.
  Thus, the Coulomb moduli and the mass parameters are on different footings.
    
  We will see below that these relations are slightly modified in the $D_n$ theory case.
  First of all, let us consider $n=2m$.
  Our claim is that the Hitchin moduli space associated with the $D_{2m}$ theory
  is the one in which the Higgs field has a singularity of order $m+1$ at $z = 0$
    \bea
    \varphi
     \sim
           dz \left( \frac{t_{m+1}}{z^{m+1}} + \ldots + \frac{t_1}{z} + \ldots \right),
           \label{Higgs1}
    \eea
  and of order $1$ at $z = z_\infty$
    \bea
    \varphi
     \sim  
           dz \left( \frac{\tilde{t}_1}{z - z_\infty} + \ldots \right).
           \label{DnHitchin}
    \eea
  More explicitly, by comparing with (\ref{Dn+1curve2}), we identify the parameters as
    \bea
    t_{m+1}
     =     \sigma_3, ~~~
    t_{m}
     =     \frac{c_1}{2} 
           \sigma_3, ~~~
    t_{m-1}
     =     \frac{1}{2} \left( c_2 - \frac{c_1^2}{4} \right)
           \sigma_3,~~ \ldots.
    \eea
  where $\sigma_3 = \diag(1, -1)/\sqrt{2}$.

  Let us first see the dimension of the moduli space $\CM_{{\rm H}}$ 
  and interpret it from the gauge theory point of view.
  The complex dimension of $\CM_{{\rm H}}$ with the required singularity structure 
  is $\dim_{\mathbb{C}} \CM_{{\rm H}}= - 6 + 2(m+1) + 2 = 2 (m - 1)$.
  Correspondingly, the complex dimension of $\CQ$ is $m - 1$.
  With $m=1$, this has to correspond to the $D_2$ theory.
  Indeed, the $D_2$ theory is simply that of the four free half-hypermultiplets
  and does not have any modulus in agreement with $\dim_{\mathbb{C}} \CQ = 0$.
  What are the $m-1$ moduli in the $D_{2m}$ theory for $m >1$?
  As discussed at the end of section \ref{subsec:Dn}, 
  this theory has $m-1$ parameters $v_a$ which are the vevs of relevant operators.
  Therefore, we identify them with the moduli parametrizing the base space $\CQ$ of the Hitchin fibration.
  
  Indeed, this can be made more concrete by comparing the spectral curve and the Seiberg-Witten curve.
  Let us write down the local behavior of the spectral curve at $z = 0$
    \bea
    \Tr \varphi^2
     \sim
           \frac{\Tr t_{m+1}^2}{z^{2m+2}} + \ldots + \frac{2\Tr t_{m+1} t_1 + \ldots}{z^{m+2}} + \ldots
    \eea
  As already noted in (\ref{quadratic}), the terms of order higher than $1/z^{m+1}$ 
  depend only on the parameters $t_i$ which specify the singularity of the Higgs field.
  The regular part of the Higgs field will enter the equation from the order $1/z^{m+1}$ on.
  In the Seiberg-Witten curve of the $D_n$ theory, the vevs of the relevant operators $v_a$ 
  will also enter from the order $1/z^{m+1}$ as in (\ref{Dn+1curve2}).
  Therefore, the geometric meaning of the parameters of the $D_{2m}$ theory now becomes clear:
  the relevant parameters $c_i$ correspond to the parameters $t_i$ 
  of the Higgs field specifying the singular part. 
  The Hitchin fibration varies holomorphically with them.
  On the other hand, the vevs of the relevant operators $v_a$ correspond 
  to the moduli of the base space of the Hitchin fibration.
  So, we see that the parameters are on different footings, as in the case with regular singularities.
    
  Let us now discuss the case $n= 2m-1$ ($m \geq 1$).
  At first sight, we encounter a contradiction because the most singular term of the spectral curve   
  is always of even degree.
  Therefore, it is impossible to describe this case from solutions of section \ref{subsec:Hitchin}.
  Notice however that 
  when we wrote down (\ref{irregularsol}) we restricted to $t_i \in \mathfrak{t}_{\mathbb{C}}$.
  Thus, we relax this condition and allow $t_{m}$ to be a nilpotent element of $G_{\mathbb{C}}$, 
  which, in the $SL(2, \mathbb{C})$ case, corresponds to $\sigma_\pm = \sigma_1 \pm i \sigma_2$.
    
  As discussed in \cite{WildWitten}, after a gauge transformation, 
  one can recover an analogous local behavior to the previous case but on the double cover of the $z$-plane,
  namely 
    \bea
    A
     =     0, ~~~~
    \phi
     =     dt \left( \frac{s_{m}}{t^{2m}} + \frac{s_{m-1}}{t^{2(m-1)}} + \ldots
         + \frac{s_1}{t^2} \right) + \ldots + c.c..
    \eea
  where $s_i \in \mathfrak{t}_{\mathbb{C}}$ and $t^2 = z$.
  The dimension of the $SL(2, \mathbb{C})$ Hitchin moduli space is $\dim_{\mathbb{C}} \CM_{{\rm H}} = 6(g-1) + 2 (m+1)$
  and, in the complex structure $I$, the Hitchin fibration varies holomorphically with respect to the parameters $s_i$.
  By going back to the $z$-coordinate, we get
    \bea
    A
     =     0, ~~~~
    \phi
     =     dz \left( \frac{s_{m}}{z^{m+1/2}} + \frac{s_{m-1}}{z^{m-1/2}} + \ldots
         + \frac{s_1}{z^{3/2}} \right) + \ldots + c.c.,
           \label{Higgs2}
    \eea
  where the fractional power indicates the presence of a cut in the $z$-plane.
    
  Now, our claim is that the $D_{2m-1}$ theory can be described by the Hitchin fibration
  with a singularity as above at $z = 0$ and a regular singularity as in (\ref{DnHitchin}) at $z = z_\infty$.
  The complex dimension of the Hitchin moduli space in this case is $-6+2(m+1) + 2 = 2(m-1)$.
  Indeed, the spectral curve near to $z=0$ is
    \bea
    \Tr \varphi^2
     \sim
           \frac{\Tr v_{m}^2}{z^{2m + 1}} + \frac{2 \Tr v_{m}v_{m-1}}{z^{2m}} + \ldots
         + \frac{2 \Tr v_{m} v_1 + \ldots}{z^{m+2}} + \ldots.
    \eea
  As in the previous case, the terms less singular than $1/z^{m+2}$ include the regular terms,
  and the dimension of the space $\CQ$ of these quadratic differentials is $m-1$.
  So, the parameters $s_i$ are related with the relevant deformation parameters $c_i$
  and the moduli of $\CQ$ corresponds to the vevs of the $m-1$ relevant operators $v_a$.
  Therefore, the geometric meaning of the gauge theory parameters is the same as in the $n=2m$ case.

\subsection{Wild quiver gauge theories and Hitchin systems}
\label{subsec:generalHitchin}
  In this subsection, we shortly consider the Hitchin moduli space corresponding to $SU(2)$ wild quiver gauge theories 
  in section \ref{subsec:generalization}.
  Associated with $\CC_{g,\ell,\{ n_\alpha \}}$ where $\alpha = 1, \ldots, k$, 
  we constructed $SU(2)$ quiver gauge theory with $k$ strongly coupled sectors $D_{n_\alpha}$.
  Here $n_\alpha$ is the degree minus $2$ of the singularity of the quadratic differential $\phi_2$
  (\ref{localSW2}).
  
  The corresponding Hitchin moduli space is formulated on a genus $g$ Riemann surface 
  with $k$ irregular and $\ell$ regular singularities.
  The singularity structure of the Higgs field is specified by (\ref{Higgs1}) when $n_\alpha = 2m$ 
  and (\ref{Higgs2}) when $n_\alpha = 2m-1$, and by (\ref{DnHitchin}) for regular singularities.
  Let us check the dimension of the Hitchin moduli space.
  As in the previous subsection, the singular behavior of the Higgs field, 
  corresponding to an irregular singularity of the $D_{n}$ theory, contributes to
  the complex dimension by $2[\frac{n+3}{2}]$.
  Moreover, each regular singularity contributes by $2$ to the dimension of the Hitchin moduli space.
  Therefore, the complex dimension of the Hitchin moduli space is 
  $\dim_{\mathbb{C}} \CM_{{\rm H}} = 6 (g - 1) + 2 \ell + 2 \sum_{\alpha = 1}^k [\frac{n_\alpha + 3}{2}]$
  and correspondingly, 
    \bea
    \dim_{\mathbb{C}} \CQ
     =     3g - 3 + \ell + \sum_{\alpha = 1}^k \left[ \frac{n_\alpha + 3}{2} \right].
    \eea
  This agrees with the counting of the moduli on the gauge theory side (\ref{numbermoduli}).

\section{Irregular conformal blocks}
\label{sec:CFT}
  In \cite{AGT}, it was proposed 
  that the instanton partition function of a weakly coupled $\CN=2$, $SU(2)$ gauge theory associated with
  a particular marking of the Riemann surface $\CC_{g,\ell}$
  can be obtained from the Virasoro conformal block on $\CC_{g,\ell}$.
  In this section we claim that 
  the partition function of the wild quiver gauge theory 
  can be obtained from irregular conformal blocks on $\CC_{g, \ell, \{ n_\alpha \}}$.
  To make this statement more precise, we have to specify what corresponds to the basic building block $\CC_{0,1,\{ n \}}$.
  It is already known that for $n = 1,2$ cases, this can be described 
  by the coherent state in the Verma module \cite{Gaiottostate}\footnote{Analogous coherent states were discussed 
  in \cite{Taki, KPPW, Wyllard, BF, WyllardALE, KMSTach, BE} 
  for various different conformal algebras. These states are also called Whittaker vectors in mathematics.}.
  Here, we need to find the generalization of this state, 
  which we will refer to as $\left| G_n \right>$ corresponding to $\CC_{0,1,\{ n \}}$.
  
  First of all, let us review the properties of the state for $n=1,2$.
  For $n=1$ the state is specified by the coherent condition:
    \bea
    L_1 \left| G_1 \right>
     =     \hat{\Lambda}^2 \left| G_1 \right>, ~~~
    L_n \left| G_1 \right>
     =     0, ~~({\rm for}~ n > 1)
           \label{Gaiotto1}
    \eea
  in the Verma module of conformal weight $\Delta$.
  Equivalently, this state can be written as \cite{MMM}
    \bea
    \left| G_1 \right>
     =     \sum_{k=0}^\infty \hat{\Lambda}^{2k} Q^{-1}_\Delta (1^k; Y) L_{-Y} \left| \Delta \right>,
           \label{Gaiotto1'}
    \eea
  where $\left| \Delta \right>$ is the primary state of weight $\Delta$ and
  $Q_{\Delta}^{-1}$ is the inverse of the Shapovalov matrix:
  $Q_{\Delta}(W;Y) = \left< \Delta \right| L_{W} L_{-Y} \left| \Delta \right>$
  and we use the notation $L_{-Y} = (L_{-1})^{m_1} (L_{-2})^{m_2} \dots$ for
  $Y = \{Y_1, Y_2, \ldots \} = [1^{m_1} 2^{m_2} \ldots]$.
  Indeed, one can show that $\left< \Delta \right| L_{Y} \left| G_1 \right> = \hat{\Lambda}^{2 k} \delta_{Y,1^k}$,
  and this implies (\ref{Gaiotto1}).
  
  The inclusion of the mass parameter, that is the state corresponding to the $D_2$ theory, is specified by
    \bea
    L_1 \left| G_2 \right>
     =     \hat{m} \hat{\Lambda} \left| G_2 \right>, ~~~
    L_2 \left| G_2 \right>
     =     \hat{\Lambda}^2 \left| G_2 \right>, ~~~
    L_n \left| G_2 \right>
     =     0, ~~({\rm for}~ n > 2) \ .
           \label{Gaiotto2}
    \eea
  This state can be written as
    \bea
    \left| G_2 \right>
     =     \sum_{k=0}^\infty \sum_{p=0} \hat{m}^{k-2p} \hat{\Lambda}^{k} 
           Q^{-1}_\Delta (2^{p} 1^{k-2p}; Y) L_{-Y} \left| \Delta \right>,
           \label{Gaiotto2'}
    \eea
  where the sum over $p$ is taken such that $k-2p$ is not negative.
  As above, one can check that 
  $\left< \Delta \right| L_{Y} \left| G_2 \right> = \hat{m}^{k-2p} \hat{\Lambda}^{k} \delta_{Y, 2^{p} 1^{k-2p}}$,
  which leads to (\ref{Gaiotto2}).
  
  By using these states, one can write the Nekrasov partition function of $SU(2)$ gauge theories with $N_f = 0, 1, 2$,
  which are the $\hat{A}_{m, n}$ theories with $m, n = 1,2$.
  For the $\hat{A}_{1,1}$ theory, it was checked in \cite{Gaiottostate} that 
  the partition function is just the norm of the simplest state:
    \bea
    Z_{Nek}^{\hat{A}_{1,1}}
     =     \left< G_1 | G_1 \right>.
           \label{AGTA11}
    \eea
  Moreover, for the $\hat{A}_{1,2}$ and $\hat{A}_{2,2}$ theories we have
  $Z_{Nek}^{\hat{A}_{1,2}} = \left< G_1 | G_2 \right>$ and $Z_{Nek}^{\hat{A}_{2,2}} = \left< G_2 | G_2 \right>$.
  The identification of the parameters is as follows.
  First of all, since the parameters in the conformal block are dimensionless,
  we have to supply a scale which we denote by $\hbar$.
  The Nekrasov deformation parameters are then identified as
    \bea
    \epsilon_1
     =     b \hbar, ~~~
    \epsilon_2
     =   - \hbar/b.
    \eea
  In other words, $\hbar^2 = - \epsilon_1 \epsilon_2$.
  The mass parameters and the vev of the scalar multiplet are identified as
  $m = \hbar \hat{m}$, $i a = \hbar \alpha$ and $\Lambda = \hbar \hat{\Lambda}$, 
  where $\alpha$ is the internal momentum and the conformal dimension is $\Delta = \frac{Q^2}{4} - \alpha^2$.
  Note that we identified the scalar multiplet vev up to an $i$ factor for later convenience.
  The relation (\ref{AGTA11}) was proved in \cite{HJS}
  by using recursion relations \cite{Zamolodchikov:1985ie, Poghossian, HJS0, FL}.
  
  In \cite{AGT}, it was found that the Seiberg-Witten curve 
  can be obtained from the classical limit 
  $\epsilon_{1,2}\to 0$
  of the vev of the energy-momentum tensor in the conformal block.
  This works also in the above examples \cite{Gaiottostate}:
    \bea
    - \epsilon_1 \epsilon_2 \frac{\left< G_1 \right| T(z) \left| G_1 \right>}{\left< G_1 | G_1 \right>}
     \rightarrow
           \frac{\Lambda^2}{z^3} + \frac{U}{z^2} + \frac{\Lambda^2}{z}
     \equiv
           \phi_2^{{\rm CFT}},
           \label{A11Tinsertion}
    \eea
  where we have taken the gauge theory limit $\epsilon_{1,2} \rightarrow 0$.
  $U$ is denoted by
    \bea
    U
     =     \lim_{\epsilon_1,\epsilon_2 \rightarrow 0} (- \epsilon_1 \epsilon_2)
           \frac{\left< G_1 \right| L_0 \left| G_1 \right>}{\left< G_1 | G_1 \right>}
     =     a^2 - \lim_{\epsilon_1,\epsilon_2 \rightarrow 0} \frac{\epsilon_1 \epsilon_2}{4} 
           \frac{\partial \ln \left< G_1 | G_1 \right>}{\partial \ln \Lambda}
     =     a^2 + \frac{1}{4} \frac{\partial F}{\partial \ln \Lambda},
    \eea
  where we have defined 
  $\left< G_1 | G_1 \right> = e^{- \frac{F}{\epsilon_1 \epsilon_2} + \ldots}$.
  This $U$ therefore coincides with the Coulomb moduli $u$ in the Seiberg-Witten curve 
  by using the Matone relation \cite{Matone}.
  Thus, (\ref{A11Tinsertion}) agrees with (\ref{curveA11}).
  Similarly, it is easy to see that
    \bea
    - \epsilon_1 \epsilon_2 \frac{\left< G_1 \right| T(z) \left| G_2 \right>}{\left< G_1 | G_2 \right>}
    &\rightarrow&
           \frac{\Lambda^2}{z^4} + \frac{m \Lambda}{z^3} + \frac{U}{z^2} + \frac{\Lambda^2}{z},
           \nonumber \\
    - \epsilon_1 \epsilon_2 \frac{\left< G_2 \right| T(z) \left| G_2 \right>}{\left< G_2 | G_2 \right>}
    &\rightarrow&
           \frac{\Lambda^2}{z^4} + \frac{m \Lambda}{z^3} + \frac{U}{z^2} + \frac{\tilde{m} \Lambda}{z} + \Lambda^2,
    \eea
  which are (\ref{curveA21}) and (\ref{curveA22}).

\subsection{Generalization of the coherent states}
  The form of the Seiberg-Witten curve and the above observations suggest that 
  the partition functions of the $\hat{A}_{m,n}$ theory and of more general wild quiver gauge theories can be obtained 
  by generalizing the above coherent state.
  However, the naive generalization like 
  $L_k \left| G_n \right> \sim \left| G_n \right>$ for $k \leq k_0$, and $L_k \left| G_n \right> =0$ for $k>k_0$
  is inconsistent with the Virasoro algebra.
  Thus we can no longer use the coherent condition to describe these theories.
  It turns out to be easier to approach the problem using the explicit expression of the state like
  (\ref{Gaiotto1'}) and (\ref{Gaiotto2'}).
  
  As a generalization of the states $\left| G_1 \right>$ and $\left| G_2 \right>$, we introduce
    \bea
    \left| G_n \right>
    &=&    \sum_{k=0}^\infty \sum_{\ell_p}
           \hat{\Lambda}^{2k/n} \prod_{i=1}^{[\frac{n}{2}]} \hat{c}_i^{\ell_{n-i}} 
           \prod_{a=1}^{[\frac{n-1}{2}]} \hat{v}_a^{\ell_a}
           Q^{-1}_\Delta (n^{\ell_n} (n-1)^{\ell_{n-1}} \cdots 2^{\ell_2} 1^{\ell_1}; Y)
           L_{-Y} \left| \Delta \right>,
           \label{gn}
    \eea
  with $\ell_1 = k - \sum_{m=2}^{n} m \ell_m$,
  which is associated with the $D_n$ theory.
  The parameters $\hat{\Lambda}$, $\hat{c}_i$ and $\hat{v}_a$ are identified with the dynamical scale of the theory
  and the parameters of the $D_n$ theory as
  $\hbar \hat{\Lambda} = \Lambda$, $\hbar^{2i/n} \hat{c}_i = c_i$ and $\hbar^{2(n-a)/n} \hat{v}_a = v_a$.
  
  Let us derive the conditions satisfied by this state. 
  It is easy to see that
    \bea
    \left< \Delta \right| L_Y \left| G_n \right>
     =     \left\{ \begin{array}{ll}
           \hat{\Lambda}^{2k/n} \prod_{i=1}^{[\frac{n}{2}]} \hat{c}_i^{\ell_{n-i}} 
           \prod_{a=1}^{[\frac{n-1}{2}]} \hat{v}_a^{\ell_a}, &\quad{\rm for}~ 
           Y = n^{\ell_n} (n-1)^{\ell_{n-1}} \cdots 2^{\ell_2} 1^{\ell_1} \quad  \\
           0 ,&\quad {\rm otherwise} 
           \end{array}
           \right.
    \eea
  which implies that
    \bea
    L_1 \left| G_n \right>
     =     \hat{\Lambda}^{\frac{2}{n}} \hat{v}_{1} \left| G_3 \right>, ~~~
    L_k \left| G_n \right>
     =     0 ~~~{\rm for}~k > n.
    \eea
  To compute the action of $L_n$, we first observe that $L_Y L_n = L_{Y'} + \ldots$, 
  where the dots denote terms involving $L_{k}$ with $k>n$ and
  $Y' = n^{\ell_n + 1} (n-1)^{\ell_{n-1}} \cdots 2^{\ell_2} 1^{\ell_1}$.
  Therefore, we obtain 
  $\left< \Delta \right| L_Y L_n \left| G_n \right> = \left< \Delta \right| L_{Y'} \left| G_n \right>
  = \hat{\Lambda}^2 \left< \Delta \right| L_Y \left| G_n \right>$
  which implies
    \bea
    L_n \left| G_n \right>
     =     \hat{\Lambda}^2 \left| G_n \right>.
    \eea
  
  We note that the state (\ref{gn}) is not an eigenfunction of $L_k$ with $1 < k < n$.
  For example, for $k=n-1$ the argument goes as follows.
  Let us observe that $L_Y L_{n-1} = L_{Y^{''}} + (2-n) \ell_1 L_{Y^{'''}} + \ldots$, 
  where again the dots denote terms involving $L_k$ with $k > n$.
  Also, $Y^{''} = n^{\ell_n} (n-1)^{\ell_{n-1} + 1} \cdots 2^{\ell_2} 1^{\ell_1}$
  and $Y^{'''} = n^{\ell_n + 1} (n-1)^{\ell_{n-1}} \cdots 2^{\ell_2} 1^{\ell_1 - 1}$.
  Therefore, we obtain
    \bea
    \left< \Delta \right| L_Y L_{n-1} \left| G_n \right>
    &=&    \hat{\Lambda}^{2(n-1)/n} \hat{c}_1 \left< \Delta \right| L_Y  \left| G_n \right>
         + (2-n) \ell_1 \hat{\Lambda}^{2(k+n-1)/n} \prod_{i=1}^{[\frac{n}{2}]} \hat{c}_i^{\ell_{n-i}} 
           \prod_{a=2}^{[\frac{n-1}{2}]} \hat{v}_a^{\ell_a}
           \hat{v}^{\ell_1 - 1}_{1}
           \nonumber \\
    &=&    \hat{\Lambda}^{2(n-1)/n} \left( \hat{c}_1 + (2 - n) \frac{\partial}{\partial \hat{v}_{1}} \right)
           \left< \Delta \right| L_Y  \left| G_n \right>,
    \eea
  which implies that
    \bea
    L_{n-1} \left| G_n \right>
     =     \hat{\Lambda}^{2(n-1)/n} \left( \hat{c}_1 + (2 - n) \frac{\partial}{\partial \hat{v}_{1}} \right)
           \left| G_n \right>.
           \label{Ln-1action}
    \eea
  The action of the other $L_k$'s can be calculated in a similar way, for example
    \bea
    L_{n-2} \left| G_n \right>
    &=&    \hat{\Lambda}^{2(n-2)/n} \left( \hat{c}_2 + (3 - n) \hat{c_1} \frac{\partial}{\partial \hat{v}_{1}}
         + \frac{(2-n)(3-n)}{2} \frac{\partial^2}{\partial \hat{v}_{1}^2}
         + (4-n) \frac{\partial}{\partial \hat{v}_{2}} \right)
           \left| G_n \right>,
           \nonumber \\
    & &    \label{Ln-2action}
    \eea
  and so on.
  A generic feature is that the action of $L_{n-k}$ starts with a linear term in the corresponding parameter 
  and the remaining terms, although involved, can be written as linear differential operators in the parameters.

\subsection{Irregular conformal block on $\CC_{g, \ell, \{ n_\alpha \}}$}
\label{subsec:conformalblock}
  Now we are ready to consider the conformal block on $\CC_{g, \ell, \{ n_\alpha \}}$ (with a particular marking),
  by using the sewing procedure.
  Without irregular punctures, the only building block corresponding to $\CC_{0,3}$ is the chiral three-point function.
  Connecting two $\CC_{0,3}$'s through a tube generates the four-point conformal block
    \bea
    \sum_{Y,W} \left< \Delta_1 \right| V_{\alpha_2}  L_{-Y} \left| \Delta \right> 
           Q_\Delta^{-1}(Y;W) 
           \left< \Delta \right| L_{W} V_{\alpha_3} \left| \Delta_4 \right>.
           \label{sewing}
    \eea
  By repeatedly applying this procedure we can in principle construct the conformal block 
  on any $\CC_{g,\ell, \{ \emptyset \}}$.
  This was proposed to be equivalent to the Nekrasov partition function 
  of a weakly coupled $\CN=2$ gauge theory with vanishing beta function, 
  associated with the same (marking of) $\CC_{g, \ell, \{ \emptyset \}}$ \cite{AGT}.
  (See \cite{BMTY, Hollands} for higher genus case).
  
  The inclusion of the second building block $\CC_{0,1,\{ n \}}$ is easily understood as follows:
  connecting two regular punctures of $\CC_{0,3}$ and $\CC_{0,1,\{ n \}}$
  can be denoted by the three-point function
    \bea
    \left< \Delta_1 \right| V_{\alpha_2} \left| G_n \right>,
    \eea
  where $\left| G_n \right>$ is the state constructed in the previous subsection.
  In the gauge theory, this denotes an $SU(2)$ gauge theory coupled with two fundamental flavors, 
  corresponding to two regular punctures,
  and one strongly coupled sector $D_n$ (denoted as $\hat{D}_n$ theory in \cite{CV}).
  Note that in the case with $n=1,2$, this corresponds to $SU(2)$ gauge theory with two and three flavors
  respectively 
  and was already analyzed in \cite{Gaiottostate}.
  By further connecting this with other building blocks as in (\ref{sewing}) 
  we obtain the generic conformal block on $\CC_{g, \ell, \{ n_\alpha \}}$.
  This describes the wild quiver gauge theory constructed in section \ref{subsec:generalization}.
  
  One exception of this construction is the case $\CC_{0,0,\{ m, n \}}$, 
  namely a sphere with two irregular punctures.
  This is simply the scalar product of the states $\left| G_{m} \right>$ and $\left| G_{n} \right>$,
  and corresponds to the $\hat{A}_{m, n}$ gauge theory.
  Namely, we conjecture that the partition function of the $\hat{A}_{m, n}$ theory 
  is the scalar product of the generalized states:
    \bea
    Z^{\hat{A}_{m,n}}_{Nek}
     =     \left< G_{m} | G_{n} \right>,
    \eea
  in the appropriate identification of the parameters.
  A weaker statement, which we will check in the present paper, is that the prepotential of the  gauge theory 
  can be obtained from 
  $F = - \lim_{\epsilon_{1,2} \rightarrow 0} (\epsilon_1 \epsilon_2) \log \left< G_m | G_n \right>$.
  
  For a check of this relation we can see that the insertion of the energy-momentum tensor
  can be written as
    \bea
    & &
    - \epsilon_1 \epsilon_2 \frac{\left< G_{m} \right| T(z) \left| G_{n} \right>}{\left< G_{m} | G_{n} \right>}
     \rightarrow
           \phi_2^{{\rm CFT}}
      =    \frac{\Lambda^2}{z^{n+2}} + \frac{\Lambda^{\frac{2n-2}{d_2}} (c_1 + s_1)}{z^{n+1}} 
         + \frac{\Lambda^{\frac{2n -4}{d_2}} (c_2 + s_2)}{z^{d_2}} + \ldots + \frac{\Lambda^{\frac{2}{n}} v_{1}}{z^3}
           \nonumber \\
    & &    ~~~~~~~~~~~~~~~~~~~~~~~~~~~~~~~~~~~~~
         + \frac{U}{z^2} + \frac{\Lambda^{\frac{2}{m}} \tilde{v}_{1}}{z} + \ldots
         + \Lambda^{\frac{2m-2}{m}} (\tilde{c}_1 + \tilde{s}_1) z^{m-3} + \Lambda^2 z^{m-2}
    \eea
  where $c_i$, $v_a$ and $\tilde{c}_i$, $\tilde{v}_a$ are the parameters 
  in the $\left| G_{m} \right>$ and $\left| G_{n} \right>$ states
  and we supplied the dimension to the parameters.
  $s_i$ and $\tilde{s}_i$ can be written in terms of 
  the derivatives of $F$ with respect to $v_a$ and $\tilde{v}_a$, 
  which comes from the derivative terms
  in the definition of the state (\ref{Ln-1action}) and (\ref{Ln-2action}).
  We defined the coefficient of the double pole as
    \bea
    U
     =     a^2 + \frac{1}{b_0} \frac{\partial F}{\partial \ln \Lambda}.
    \eea
  where $b_0$ is the one-loop beta function coefficient of the $\hat{A}_{m,n}$ theory.
  This shows that $\phi_2^{{\rm CFT}}$ has a similar structure as that of $\phi_2$.
  Let us below consider a few example more explicitly and check the agreement with the gauge theory.

\subsubsection*{Irregular conformal block for the $\hat{A}_{1,3}$ theory}
  The state $\left| G_3 \right>$ is given by
    \bea
    \left| G_3 \right>
     =     \sum_{k=0}^\infty \sum_{p, q} \hat{\Lambda}^{2k/3} \hat{c}_1^{q} \hat{v}_1^{k-3p-2q}
           Q^{-1}_\Delta (3^{p} 2^{q} 1^{k-3p-2q}; Y) L_{-Y} \left| \Delta \right>,
    \eea
  which satisfies $\left< \Delta \right| L_Y \left| G_3 \right> = \hat{\Lambda}^{2k/3} \hat{c}_1^q \hat{v}_1^{k-3p-2q}$
  for $Y=3^p2^q1^{k-3p-2q}$.
  From the general argument above, this state is also specified by
    \bea
    L_1 \left| G_3 \right>
     =     \hat{\Lambda}^{\frac{2}{3}} \hat{v}_1 \left| G_3 \right>, ~~~
    L_2 \left| G_3 \right>
     =     \hat{\Lambda}^{\frac{4}{3}} \left( \hat{c}_1 - \frac{\partial}{\partial \hat{v}_1} \right) \left| G_3 \right>, 
           ~~~
    L_3 \left| G_3 \right>
     =     \hat{\Lambda}^2 \left| G_3 \right>,
           \label{LactionG3}
    \eea
  and $L_k \left| G_3 \right> = 0$ for $k > 3$.
  We will consider the scalar product $\left< G_1 | G_3 \right>$.
  
  The insertion of the energy-momentum tensor can be written as
    \bea
    - \epsilon_1 \epsilon_2 \frac{\left< G_1 \right| T(z) \left| G_3 \right>}{\left< G_1 | G_3 \right>}
     \rightarrow
           \phi_2^{{\rm CFT}}
      =    \frac{\Lambda^2}{z^5} + \frac{\Lambda^{4/3} (c_1 + s)}{z^4} + \frac{\Lambda^{2/3} v_1}{z^3}
         + \frac{U}{z^2} + \frac{\Lambda^2}{z},
           \label{A31T}
    \eea
  where we supplied the dimensions to the parameters and introduced 
    \bea
    s
     =   - \frac{\partial F}{\partial v_1}.
           \label{Us}
    \eea
  Note that $s$ comes from the derivative term in the $L_2$ action on $\left| G_3 \right> $ (\ref{LactionG3}).
  
  Let us then compute the scalar product explicitly.
  By expanding in $\Lambda$, this is written as
    \bea
    \left< G_1 | G_3 \right>
     =     \sum_{k=0} \Lambda^{8k/3} \CB_k,
    \eea
  where $\Lambda = \hbar \hat{\Lambda}$, $\CB_0 = 1$ and the lowest orders are 
    \bea
    \CB_1
     =     \frac{v_1}{2 (\epsilon_1 \epsilon_2)^2 \Delta}, ~~~
    \CB_2
     =     \frac{(8 \Delta + c) v_1^2 - 12 \Delta c_1}{4 (\epsilon_1 \epsilon_2)^4 
           \Delta (2 \Delta (8 \Delta - 5) + (1 + 2 \Delta)c)},~~ \ldots.
    \eea
  Then, it is easy to get
    \bea
    F
     =     \frac{v_1}{2a^2} \Lambda^{8/3} + \frac{5v_1^2 - 12c_1 a^2}{64 a^6} \Lambda^{16/3}
         + \frac{9 v_1^3 - 28 c_1 v_1 a^2 + 32 a^4}{192 a^{10}} \Lambda^{8} + \ldots
    \eea
  One may think that this might be equivalent to the prepotential of the $\hat{A}_{1,3}$ theory.
  However, there is a subtlety associated to the presence of $s$ in (\ref{A31T}).
  As we will see in detail in a moment, this provides a redefinition of the $c_1$ modulus
  of the CFT curve which gives back the $u$ modulus of the Seiberg-Witten curve.
  We will postpone the discussion on the agreement with the prepotential to the next subsection.
  
  At this stage, we can however check at least the equivalence between $\phi_2$ and $\phi_2^{{\rm CFT}}$.
  In order to get the same expression, we have to identify the parameter $c_1$ in the gauge theory with $c_1 + s$
  in (\ref{A31T}).
  A subtlety here is that $s$ is an infinite series in $\Lambda$.
  However, we can do the identification order by order in the $\Lambda$ expansion.
  Then $U$, which depends on $c_1$, should be considered under this identification
    \bea
    U
    &=&    a^2 + \frac{3}{8} \frac{\partial F}{\partial \ln \Lambda} \Bigg|_{c_1 \rightarrow c_1 - s}
           \nonumber \\
    &=&    a^2 + \frac{v_1}{2a^2} \Lambda^{8/3} + \frac{5 v_1^2 - 12 c_1 a^2}{32 a^{6}} \Lambda^{16/3}
         + \frac{9 v_1^3 - 28 c_1 v_1 a^2 + 20 a^4}{64a^{10}} \Lambda^8
         + \ldots.
    \eea
  We can see that this agrees with the $u$ modulus (\ref{uA13}) calculated in the Appendix \ref{sec:period}
from the Seiberg-Witten curve.

\subsubsection*{Irregular conformal block for the $\hat{A}_{1,4}$ theory}
  Let us next consider the state $\left| G_4 \right>$ which is characterized by
    \bea
    L_1 \left| G_4 \right>
    &=&    \hat{\Lambda}^{\frac{1}{2}} \hat{v}_1 \left| G_4 \right>, ~~~
    L_2 \left| G_4 \right>
     =     \hat{\Lambda} \left( \hat{c}_2 - \hat{c_1} \frac{\partial}{\partial \hat{v}_{1}}
         + \frac{\partial^2}{\partial \hat{v}_{1}^2} \right) \left| G_4 \right>,
           \nonumber \\
    L_3 \left| G_4 \right>
    &=&    \hat{\Lambda}^{\frac{3}{2}} \left( \hat{c}_1 - 2 \frac{\partial}{\partial \hat{v}_1} \right) \left| G_4 \right>,
           ~~~
    L_4 \left| G_4 \right>
     =     \hat{\Lambda}^2 \left| G_4 \right>,~~~
    L_k \left| G_4 \right>
     =     0, ~~(k>4)
           \label{Gaiotto4}
    \eea
  We consider the scalar product $\left< G_1 | G_4 \right>$.
  The insertion of the energy-momentum tensor and the limit $\epsilon_{1,2} \rightarrow 0$ lead to
    \bea
    \label{phi-CFT}
    \phi_2^{{\rm CFT}}
     =     \frac{\Lambda^2}{z^6} + \frac{\Lambda^{3/2} (c_1 + s_1)}{z^5} + \frac{\Lambda (c_2 + s_2)}{z^4}
         + \frac{\Lambda^{1/2} v_1}{z^3} + \frac{U}{z^2} + \frac{\Lambda^2}{z},
    \eea
  where 
    \bea
    s_1
     =   - 2 \frac{\partial F}{\partial v_1}, ~~~~
    s_2
     =   - c_1 \frac{\partial F}{\partial v_1} - \left( \frac{\partial F}{\partial v_1} \right)^2.
           \label{s1s2}
    \eea
  These are obtained from the derivative terms in (\ref{Gaiotto4}). As explained for the previous
  case, in order to compare with the Seiberg-Witten curve 
  we have to identify the gauge theory parameters $c_i$ ($i=1,2$)  
  with $c_i + s_i$ appearing in (\ref{phi-CFT}).
  The function $F$ can be calculated to be 
    \bea
    F
     =     \frac{v_1}{2 a^2} \Lambda^{5/2} + \frac{5 v_1^2 - 12 c_2 a^2}{64 a^{6}} \Lambda^5
         + \frac{9 v_1^3 - 28 c_2 v_1 a^2 + 32 c_1 a^4}{192 a^{10}} \Lambda^{15/2}
         + \ldots.
    \eea
  As in the previous case, the value of $U$ after the shifting $c_i \rightarrow c_i - s_i$ ($i=1,2$)
  can be checked to agree with the $u$ modulus (\ref{uA14}) calculated from the gauge theory.
    
\subsubsection*{Irregular conformal block for the $\hat{A}_{2,3}$ theory}
  As a last example, we consider $\left< G_2 | G_3 \right>$
  corresponding to the $\hat{A}_{2,3}$ theory.
  As before, the energy-momentum tensor insertion in the $\epsilon_{1,2}\to 0$ limit lead to
    \bea
    \phi_2^{{\rm CFT}}
     =     \frac{\Lambda^2}{z^5} + \frac{\Lambda^{4/3} (c_1 + s)}{z^4} + \frac{\Lambda^{2/3} v_1}{z^3}
         + \frac{U}{z^2} + \frac{\Lambda m}{z} + \Lambda^2,
    \eea
  where $s$ is expressed by (\ref{Us}), and
  $F$ can be calculated as
    \bea
    F
    &=&    \frac{m v_1}{2 a^2} \Lambda^{5/3}
         + \frac{5m^2 v_1^2 - 12 (m^2 c_1 + v_1^2) a^2 + 16c_1 a^4}{64a^6} \Lambda^{10/3}
           \nonumber \\
    & &    ~~~~~
         + \frac{m (9 m^2 v_1^3 - 28 (m^2 c_1 v_1 + v_1^3) a^2 + (80 c_1 v_1 + 32 m^2) a^4 - 64 a^6)}{192 a^{10}} 
           \Lambda^5 + \ldots
    \eea
  Again by taking the existence of $s$ in $\phi_2^{{\rm CFT}}$ into account,
  we find that $U$ agrees with the gauge theory result (\ref{uA14}).

\subsection{Insertion of degenerate field}
\label{subsec:degenerate}
  In this subsection, we consider the insertion of a degenerate field in the conformal blocks.
  We concentrate on the degenerate field $\Phi_{1,2}$, 
  which is the operator with Liouville momentum $-\frac{1}{2b}$ 
  (and thus the dimension $\Delta_{1,2} = - \frac{1}{2} - \frac{3}{4 b^2}$) 
  and define the conformal block with the additional degenerate field as $\Psi(z)$, 
  e.g., in the case of the scalar product of the states that we considered in the previous subsection we define
    \bea
    \Psi(z)
     =     \left< G_{m} \right| \Phi_{1,2}(z) \left| G_{n} \right>.
           \label{Psi}
    \eea
  We will obtain below the second order differential equation satisfied by (\ref{Psi}) which follows from 
  the null field condition $(b^{-2} L_{-2} + (L_{-1})^2) \Phi_{1,2} (z) = 0$ \cite{BPZ}.
  The equations for the case with $m = n = 1$ and with $m, n \leq 2$ were calculated in 
  \cite{AY} and \cite{MT, Awata:2010bz} respectively.
  We will also consider the monodromies of $\Psi$ along some non-contractible cycles of the Riemann surface.
  As in \cite{AGGTV}, this leads, in the $\epsilon_{1,2} \rightarrow 0$ limit, to
  the special geometry relation identified with the Seiberg-Witten one.
  By using this idea, we will calculate the prepotential of the gauge theory 
  from the CFT analysis performed in the previous subsection.
  
  In \cite{AGGTV}, it was claimed that the insertion of the degenerate field corresponds 
  to the Nekrasov partition function in presence of a surface operator.
  This was checked and analyzed in \cite{KPW, DGH, MT, Awata:2010bz, BTZ}.
  It is natural to think that $\Psi$ here describes the surface operator insertion in the wild quiver gauge theory.
  
  First of all, we note that in the limit $\epsilon_{1, 2} \rightarrow 0$, the semiclassical expansion of (\ref{Psi})
  dictates the
 dependence on $z$ to start from the subleading order in $\hbar$ as
    \bea
    \Psi
     =     \exp \left( - \frac{1}{\epsilon_1 \epsilon_2} (F + \frac{\hbar}{b} \CW(z) + \CO(\hbar^2)) \right),
           \label{Psiepsilon2}
    \eea
  where the first term is the leading term in the scalar product that we computed in the previous subsection.
  
  Let us then derive the differential equation for $\Psi(z)$.
  While this can be obtained in any irregular conformal block which includes several generalized states,
  we focus here on the case of (\ref{Psi}).
  Let $\Delta=\Delta(\alpha -b/4)$ and $\Delta^{\prime}=\Delta(\alpha + b/4) $
  be the conformal dimensions of the level zero parts of $\left| G_{m} \right>$ and $\left| G_{n} \right>$,
  in accordance with the fusion rule.
  Then, what we need to calculate is $\left< G_{m} \right| L_{-2} \Phi_{1,2}(z) \left| G_{n} \right>$.
  In order to do that, we consider the insertion of the energy momentum tensor:
    \bea
    & &    
    \left< G_{m} \right| T(w) \Phi_{1,2}(z) \left| G_{n} \right>
           \nonumber \\
    & &    
     =     \sum_{n=0}^\infty \frac{1}{w^{n+2}} \left< G_{m} \right| [ L_n, \Phi_{1,2}(z)] \left| G_{n} \right>
         + \frac{1}{w^2} \left< G_{m} \right| \Phi_{1,2}(z) L_0 \left| G_{n} \right>
         - \frac{1}{\epsilon_1 \epsilon_2} \hat{\phi}_2^{{\rm CFT}} 
           \left< G_{m} \right| \Phi_{1,2}(z) \left| G_{n} \right>
           \nonumber \\
    & &    
     =     \Bigg[ \frac{z}{w(w-z)} \frac{\partial}{\partial z} + \frac{\Delta_{1,2}}{(w-z)^2}
         - \frac{1}{\epsilon_1 \epsilon_2} \hat{\phi}_2^{{\rm CFT}}
           \nonumber \\
    & &    ~~~~~~~~~~
         + \frac{2}{b_0 w^2} \left( - \frac{z}{m} \frac{\partial}{\partial z}
         - \frac{\Delta_{1,2}}{m} + \frac{1}{2}\frac{\partial}{\partial \ln \Lambda}
         + \frac{m \Delta + n \Delta'}{m n} \right) \Bigg] \Psi(z),
           \label{TPhi}
    \eea
  where we used 
  $[L_n, \Phi_{1,2}(z)] = \left( z^{n+1} \partial_z + (n+1) z^n \Delta_{1,2} \right) \Phi_{1,2}$
  and 
    \bea
    \frac{\partial}{\partial \ln \Lambda} \Psi(z)
     =     \frac{2}{m} \left< G_{m} \right| [L_0, \Phi_{1,2}(z)] \left| G_{n} \right>
         + b_0 \left< G_{m} \right| \Phi_{1,2}(z) L_0 \left| G_{n} \right>
         - 2\left(\frac{\Delta'}{m} + \frac{\Delta}{n} \right) \Psi(z),
           \nonumber 
    \eea
  in order to rewrite $\left< G_{m} \right| \Phi_{1,2}(z) L_0 \left| G_{n} \right>$ in the second line.
  Moreover, we defined $\hat{\phi}_2^{{\rm CFT}}$ as
    \bea
    \hat{\phi}_2^{{\rm CFT}} 
     =     \phi_2^{{\rm CFT}} - \frac{U}{z^2}.
    \eea
  By reading off the coefficients of $(w-z)^{0}$ in (\ref{TPhi}), 
  we get $\left< G_{m} \right| L_{-2} \Phi_{1,2}(z) \left| G_{n} \right> = \hat{L} \Psi(z)$ where
    \bea
    \hat{L}
    &=&  - \frac{1}{\epsilon_1 \epsilon_2} \hat{\phi}_2^{{\rm CFT}}
         - \frac{1}{z^2} \left(1 + \frac{2}{b_0 m} \right) z \frac{\partial}{\partial z}
         + \frac{2}{b_0 z^2} \left( - \frac{\Delta_{1,2}}{m} + \frac{1}{2}\frac{\partial}{\partial \ln \Lambda}
         + \frac{m \Delta + n \Delta'}{m n} \right).
    \eea
  Therefore, the differential equation is
    \bea
    \left( b^2 \frac{\partial^2}{\partial z^2} + \hat{L} \right) \Psi(z)
     =     0.
    \eea
  As discussed in \cite{Teschner, MT, MMMint, BMTintegral} (see also \cite{Tai, Piatek}),
  this equation, in the $\epsilon_2\to 0$ limit, is related to the quantization of the corresponding integrable system, 
  namely the Hitchin system with wild ramification.
  The quantization of the related Gaudin model with irregular singularity was discussed 
  e.g. in \cite{FFR, Frenkel, FFT}.
  It would be interesting to study this direction further.
    
  Here, we are interested in the limit $\epsilon_{1,2} \rightarrow 0$. 
  It follows from the expansion $- \epsilon_1 \epsilon_2 \Delta = a^2 + \ldots$ and a similar one for $\Delta'$, that
    \bea
    \lim_{\epsilon_{1,2} \rightarrow 0} (- \epsilon_1 \epsilon_2) \hat{L} \Psi
     =     \phi_2^{{\rm CFT}} \Psi.
    \eea
  Therefore, we finally obtain in the scaling limit
    \bea
    \left( (b \hbar)^2 \frac{\partial^2}{\partial z^2} + \phi_2^{{\rm CFT}} \right) \Psi
     =     0.
    \eea
  By formally solving this, we get
    \bea
    \CW(z)
     =   \pm i \int^z \sqrt{\phi_2^{{\rm CFT}}} dz',
    \eea
  where $\CW(z)$ was defined (\ref{Psiepsilon2}).
  The $\pm$ sign reflects the two-fold degeneracy of the solution to the quadratic differential equation.
  In what follows, we consider the case with the plus sign.
  
  As found in \cite{AGGTV}, the monodromies of the conformal block 
  with a degenerate field insertion along the $A$- and $B$-cycles correspond 
  to Wilson and t' Hooft loop operators on the surface operator in the gauge theory.
  In \cite{AGGTV}, these monodromies have been calculated in the conformal field theory:
    \bea
    \Psi (a, z + A)
     =     \exp \left(\frac{2 \pi a}{\hbar b} \right) \Psi(a, z), ~~~~~
    \Psi (a, z + B)
     =     \Psi (a + \frac{i \hbar}{2 b}, z),
           \label{monodromyPsi}
    \eea
  where $\Psi(z + A({\rm or~}B))$ denotes the monodromy along the $A$(or $B$)-cycle.
  Since $\Psi$ is expanded in $\hbar$ as (\ref{Psiepsilon2}), after the semi-classical expansion 
  we obtain
    \bea
    \oint_A \sqrt{\phi_2^{{\rm CFT}}} dz
     =     2 \pi i a, 
           ~~~
    \oint_B \sqrt{\phi_2^{{\rm CFT}}} dz
     =     \frac{i}{2} \frac{\partial F}{\partial a}.
           \label{monodromy}
    \eea
  Note that we have already checked that $U$ in the integrand 
  can be identified with the Coulomb modulus $u$ computed from the $A$-cycle integral 
  in the $\hat{A}_{1,3}$, $\hat{A}_{2,3}$ and $\hat{A}_{1,4}$ theories.
  We expect that this result is generic for all conformal blocks involving irregular states $\left| G_n \right>$
  and the corresponding wild quiver gauge theories.
  Since the integrand is the Seiberg-Witten differential, 
  the result of the $B$-cycle integral is the same too.
  
  However, from the conformal field theory side, we do not need to calculate the $B$-cycle integral,
  since it can be obtained directly from the derivative of $F$.
  The final caution is the shift in the $c_i$ parameters found in the previous subsection.
  Indeed, only after taking into account this shift, the B-integral matches the computation from $F$, namely
  $\frac{\partial F}{\partial a}|_{c_i \rightarrow c_i - s_i}$
  agrees with the $B$-cycle integral of the Seiberg-Witten differential.
  By using the definition (\ref{SWrelation}), the prepotential $\CF$ is obtained as the primitive function in $a$ of
  $\frac{\partial F}{\partial a}|_{c_i \rightarrow c_i - s_i}$.
  
  E.g., from the scalar product $\left< G_1 | G_3 \right>$, we get
    \bea
    \CF_{\hat{A}_{1,3}}
     =     \frac{8 a^2}{3} \ln \Lambda
         + \frac{v_1}{2a^2} \Lambda^{8/3} + \frac{5 v_1^2 - 12 c_1 a^2}{64 a^{6}} \Lambda^{16/3}
         + \frac{9 v_1^3 - 28 c_1 v_1 a^2 + 20 a^4}{192 a^{10}} \Lambda^8 + \ldots
    \eea
  while from $\left< G_1 | G_4 \right>$ and $\left< G_2 | G_3 \right>$, we get
    \bea
    \CF_{\hat{A}_{1,4}}
    &=&    \frac{5 a^2}{2} \ln \Lambda
         + \frac{v_1}{2a^2} \Lambda^{5/2} + \frac{5 v_1^2 - 12 c_2 a^2}{64 a^6} \Lambda^5
         + \frac{9 v_1^3 - 28 c_2 v_1 a^2 + 20 c_1 a^4}{192 a^{10}} \Lambda^{15/2} + \ldots,
    \eea
  and
    \bea
    \CF_{\hat{A}_{2,3}}
    &=&    \frac{5 a^2}{3} \ln \Lambda
         + \frac{m v_1}{2 a^2} \Lambda^{5/3}
         + \frac{5 m^2 v_1^2 - 12 (v_1^2 + m^2 c_1) a^2 + 16 c_1 a^4}{64 a^6} \Lambda^{10/3}
           \nonumber \\
    & &    ~~
         + \frac{m( 9 m^2 v_1^3 - 28 (v_1^3 + m^2 c_1 v_1) a^2 + 20( m^2 + 4 c_1 v_1) a^4 - 52 a^6)}{192 a^{10}} 
           \Lambda^5
         + \ldots.
    \eea
  respectively.
  
  Note that the monodromies found in \cite{AGGTV} are valid for the regular conformal blocks
  corresponding to the $\CN=2$ superconformal gauge theories.
  The monodromies of the irregular conformal block have not been calculated yet.
  However, in some cases we can verify this:
  for the $\hat{A}_{1,3}$ theory which is obtained from the $SU(2) \times SU(2)$ superconformal theory,
  the corresponding conformal block might be also obtained from the five-point regular conformal block.
  The limit which one takes to get the $\hat{A}_{1,3}$ theory does not affect the monodromies
  and therefore (\ref{monodromy}) is correct in this case.

\section{Conclusions and discussions}
\label{sec:conclusions}
  In this paper we proposed a quantitative approach to calculate the full prepotential in the $\Omega$-background 
  of $SU(2)$ wild quiver gauge theories coupled to nontrivial SCFTs via the AGT correspondence.
  
  It would be interesting to generalize the construction of wild quivers to the higher rank case.
  Indeed, when we consider the $A_{N-1}$ $(2,0)$ theory on a Riemann surface, 
  various types of singularities, labeled by Young diagrams, can be allowed 
  \cite{Gaiotto,IMO, KMST, ND3, CD, DP, Tachikawa}.
  These corresponds to $\CN=2$ quiver gauge theories with vanishing beta function coefficients.
  More in general, it is possible to consider irregular singularities also in the higher rank case. 
  These correspond to asymptotically free gauge theories, as exemplified in \cite{GMN, NX2} and,
  in the $A_1$ case, reduce to the $D_1$ and the $D_2$ type singularities studied in our paper.
  Therefore, our results suggest to investigate more general singularities 
  and the corresponding irregular conformal blocks 
  which should give a generalization of the one found in \cite{Taki} for the $SU(3)$/$\CW_3$ case.
  
  We observe that it would be useful to gain insight in the CFT on a more direct and geometrical construction 
  of the coherent state and its generalizations that we discussed in section \ref{sec:CFT}. 
  In the specific case of the Liouville theory, the operator creating an irregular puncture is
  naturally induced by the boundary condition at the insertion point resulting by the solution of the
  classical Liouville field generating higher order singularities in the classical stress-energy tensor.
  The very definition of the operator is anyway independent on the specific CFT at hand
  and having a geometric counterpart of the state building recipe (\ref{gn}) would be interesting.
  Also the role of these states in the matrix model approach to AGT correspondence 
  \cite{DV, EM, MMS} should be clarified.
  
  Our construction of the irregular conformal blocks is shown to be strictly related to 
  Hitchin systems with wild ramification and provides a scheme to quantize them which should be further developed.
  This should be obtained by analyzing the irregular conformal block 
  in the $\epsilon_2 \rightarrow 0$ limit \cite{NS1,NS2}.
  
  Last but not least our results pave the way towards a topological string interpretation 
  of the strongly couples systems which would be very interesting to analyze.
  A useful tool in this context would be the study of the generalized holomorphic anomaly equation, 
  as done for example in \cite{HKK}.

\section*{Acknowledgments}
We would like to thank 
E.~Frenkel for useful comments on the Gaudin model,
and F.~Yagi
for discussions on the calculation of the contour integrals of the Seiberg-Witten differential.
We would like to thank 
T.~Eguchi,
B.~Feigin,
G.~Giribet, 
K.~Hori,
K.~Hosomichi, 
S.~Pasquetti,
V.~Roubtsov,
S.~Sugimoto and
M.~Taki for comments and discussions.
K.M. would like to thank IPMU and KITP for hospitality during the course of this project.
G.B. is supported in part by the MIUR-PRIN contract 2009-KHZKRX. 
G.B. and K.M. are partially supported by the INFN project TV12. 
A.T. is partially supported by PRIN ``Geometria delle variet\'a algebriche e loro spazi di moduli'' and the INFN project PI14
``Nonperturbative dynamics of gauge theories''.


\appendix

\section*{Appendix}

\section{Computation of $u(a)$}
\label{sec:period}
  In this appendix we calculate the $A$-cycle integral of the Seiberg-Witten differential.
  As seen in section \ref{subsec:degenerate}, it is enough to compute it
  in order to check the correspondence with the conformal block.
  While the way which will be explained here can be applied to the generic $\hat{A}_{m,n}$ case,
  we mainly consider the $\hat{A}_{1,3}$ theory for illustration.
  We also give the relevant results for the $\hat{A}_{1,4}$ and $\hat{A}_{2,3}$ theories.
  
  Let us analyze the curve of the $\hat{A}_{1,3}$ theory
  which is $x^2 = \phi_2$ where
    \bea
    \phi_2
     =     \frac{\Lambda^2}{z^5} + \frac{\Lambda^{\frac{4}{3}} c_1}{z^4}
         + \frac{\Lambda^{\frac{2}{3}} v_1}{z^3} + \frac{u}{z^2} + \frac{\Lambda^2}{z}
     =     \frac{\Lambda^2}{z^5} P_4(z).
    \eea
  The corresponding Seiberg-Witten differential is $\lambda = x dz$.
  We want to calculate the $A$-cycle integral when the dynamical scale $\Lambda$ is very small.
  This corresponds to the classical limit.
  In order to do that, we have to specify the $A$-cycle of the curve.
  As seen in section \ref{subsec:Amn}, the branch points are at 
  the roots of $P_4(z)$ and $z = 0, \infty$.
  Among the four roots of $P_4$, one of them, say $a_1$, scales as $\Lambda^{-2}$ and the others as $\Lambda^{2/3}$.
  Therefore, in the classical limit $\Lambda \rightarrow 0$, 
  the root $a_1$ collapses to infinity, and the others collapse to $z=0$.
  Thus, it is natural to take the $A$-cycle as the contour around the cut between $a_1$ and infinity.
  Then it is possible to deform the contour to the one around $z=0$ with radius $r \simeq \CO(\Lambda^0)$.
  Let this contour be $C$.
  
  Now we consider the Seiberg-Witten relation
    \bea
    2 \pi i a 
     =     \oint_A \lambda, ~~~~
    2 \pi i a_D
     =     \oint_B \lambda,
           \label{SWrelation}
    \eea
  where the prepotential is given by
    \bea
    a_D
     =     \frac{1}{4 \pi} \frac{\partial \CF}{\partial a}.
    \eea
  It follows from the observation above that the $A$-cycle integral of the differential can be expanded as
    \bea
    2 \pi i a
     =     \oint_A \lambda
     =     \oint_C \frac{\sqrt{u}}{z} \left( 1 + \frac{X}{2 u} - \frac{X^2}{8 u^2} + \ldots \right),
    \eea
  where 
    \bea
    X
     =     \frac{\Lambda^2}{z^3} + \frac{\Lambda^{\frac{4}{3}} c_1}{z^2}
         + \frac{\Lambda^{\frac{2}{3}} v_1}{z} + \Lambda^2 z
    \eea
  Note that this expansion is valid for our choice of the contour $C$.
  Since the integrand has a pole only at $z  = 0$, 
  what one has to do is to find out the coefficient in $z^{-1}$ in each order in the expansion in $\Lambda$.
  This gives the result:
    \bea
    a 
     =     \sqrt{u} \left( 1 - \frac{v_1}{4 u^2} \Lambda^{8/3} + \frac{12 c_1 u - 15 v_1^2}{64 u^4} \Lambda^{16/3}
         - \frac{40 u^2 - 140 c_1 v_1 u + 105 v_1^3}{256 u^6} \Lambda^8
         + \ldots \right).
    \eea
  By inverting this equation, we obtain
    \bea
    u
     =     a^2 + \frac{v_1}{2 a^2} \Lambda^{8/3} + \frac{5 v_1^2 - 12 c_1 a^2}{32a^6} \Lambda^{16/3} 
         + \frac{9 v_1^3 - 28 c_1 v_1 a^2 + 20 a^4}{64a^{10}} \Lambda^8 
         + \ldots.
           \label{uA13}
    \eea
  This agrees with $U$ calculated from $\left< G_1 | G_3 \right>$ in section \ref{subsec:conformalblock}.
    
  In the same way, we can calculate the $u$'s of the $\hat{A}_{1,4}$ and $\hat{A}_{2,3}$ theories.
  The results are
    \bea
    u_{\hat{A}_{1,4}}
    &=&    a^2 + \frac{v_1}{2 a^2} \Lambda^{5/2} + \frac{5 v_1^2 - 12 a^2 c_2}{32 a^6} \Lambda^5
         + \frac{9 v_1^3 - 28 c_2 v_1 a^2 + 20 c_1 a^4}{64 a^{10}} \Lambda^{15/2} + \ldots,
           \nonumber \\
    u_{\hat{A}_{2,3}}
    &=&    a^2 + \frac{m v_1}{2a^2} \Lambda^{5/3}
         + \frac{5 m^2 v_1^2 - 12 (v_1^2 + m^2 c_1) a^2 + 16 c_1 a^4}{32a^6} \Lambda^{10/3}
           \nonumber \\
    & &    ~~~~
         + \frac{m (9 m^2 v_1^3 - 28 (v_1^3 + m^2 c_1 v_1) a^2 + 20 (m^2 + 4 c_1 v_1) a^4 - 48 a^6}{64a^{10}} \Lambda^5
         + \ldots.
           \label{uA14}
    \eea
  These agree with the $U$'s computed from $\left< G_1 | G_4 \right>$ and $\left< G_2 | G_3 \right>$
  respectively.



\begin{thebibliography}{99}
\setlength{\itemsep}{-3pt}

\bibitem{AGT}
  L.~F.~Alday, D.~Gaiotto and Y.~Tachikawa,
  ``Liouville Correlation Functions from Four-dimensional Gauge Theories,''
  Lett.\ Math.\ Phys.\  {\bf 91}, 167 (2010)
  [arXiv:0906.3219 [hep-th]].

\bibitem{GMN}
  D.~Gaiotto, G.~W.~Moore and A.~Neitzke,
  ``Wall-crossing, Hitchin Systems, and the WKB Approximation,''
  arXiv:0907.3987 [hep-th].

\bibitem{BT} 
  G.~Bonelli and A.~Tanzini,
  ``Hitchin systems, N=2 gauge theories and W-gravity,''
  Phys.\ Lett.\ B\ {\bf 691}, 111  (2010)
  [arXiv:0909.4031 [hep-th]].

\bibitem{AD} 
  P.~C.~Argyres and M.~R.~Douglas,
  ``New phenomena in SU(3) supersymmetric gauge theory,''
  Nucl.\ Phys.\ B\ {\bf 448}, 93  (1995)
  [hep-th/9505062].
  
\bibitem{MN} 
  J.~A.~Minahan and D.~Nemeschansky,
  ``An N=2 superconformal fixed point with E(6) global symmetry,''
  Nucl.\ Phys.\ B\ {\bf 482}, 142  (1996)
  [hep-th/9608047];
%
  ``Superconformal fixed points with E(n) global symmetry,''
  Nucl.\ Phys.\ B\ {\bf 489}, 24  (1997)
  [hep-th/9610076].
  
\bibitem{AS} 
  P.~C.~Argyres and N.~Seiberg,
  ``S-duality in N=2 supersymmetric gauge theories,''
  JHEP\ {\bf 0712}, 088  (2007)
  [arXiv:0711.0054 [hep-th]].

\bibitem{Gaiotto}
  D.~Gaiotto,
  ``N=2 dualities,''
  arXiv:0904.2715 [hep-th].
  
\bibitem{CV}
  S.~Cecotti and C.~Vafa,
  ``Classification of complete N=2 supersymmetric theories in 4 dimensions,''
  arXiv:1103.5832 [hep-th].
  
\bibitem{Nek} 
  N.~A.~Nekrasov,
  ``Seiberg-Witten prepotential from instanton counting,''
  Adv.\ Theor.\ Math.\ Phys.\ \ {\bf 7}, 831  (2004)
  [hep-th/0206161].

\bibitem{EHIY}
  T.~Eguchi, K.~Hori, K.~Ito and S.~K.~Yang,
  ``Study of N=2 superconformal field theories in four-dimensions,''
  Nucl.\ Phys.\  B {\bf 471}, 430 (1996)
  [arXiv:hep-th/9603002].

\bibitem{Gaiottostate}
  D.~Gaiotto,
  ``Asymptotically free N=2 theories and irregular conformal blocks,''
  arXiv:0908.0307 [hep-th].





  
\bibitem{CNV}
  S.~Cecotti, A.~Neitzke, C.~Vafa,
  ``R-Twisting and 4d/2d Correspondences,''
  [arXiv:1006.3435 [hep-th]].
  
\bibitem{ARSW}
  P.~C.~Argyres, M.~Ronen Plesser, N.~Seiberg and E.~Witten,
  ``New N=2 superconformal field theories in four-dimensions,''
  Nucl.\ Phys.\  B {\bf 461}, 71 (1996)
  [arXiv:hep-th/9511154].

\bibitem{SW1} 
  N.~Seiberg and E.~Witten,
  ``Electric - magnetic duality, monopole condensation, and confinement in N=2 supersymmetric Yang-Mills theory,''
  Nucl.\ Phys.\ B\ {\bf 426}, 19  (1994)
  [Erratum-ibid.\ B\ {\bf 430}, 485  (1994)]
  [hep-th/9407087].

\bibitem{SW2} 
  N.~Seiberg and E.~Witten,
  ``Monopoles, duality and chiral symmetry breaking in N=2 supersymmetric QCD,''
  Nucl.\ Phys.\ B\ {\bf 431}, 484  (1994)
  [hep-th/9408099].

\bibitem{APS} 
  P.~C.~Argyres, M.~R.~Plesser and A.~D.~Shapere,
  ``The Coulomb phase of N=2 supersymmetric QCD,''
  Phys.\ Rev.\ Lett.\ \ {\bf 75}, 1699  (1995)
  [arXiv:hep-th/9505100 [hep-th]].

\bibitem{HO} 
  A.~Hanany and Y.~Oz,
  ``On the quantum moduli space of vacua of N=2 supersymmetric SU(N(c)) gauge theories,''
  Nucl.\ Phys.\ B\ {\bf 452}, 283  (1995)
  [hep-th/9505075].

\bibitem{GST}
  D.~Gaiotto, N.~Seiberg and Y.~Tachikawa,
  ``Comments on scaling limits of 4d N=2 theories,''
  JHEP {\bf 1101}, 078 (2011)
  [arXiv:1011.4568 [hep-th]].
  
\bibitem{AT}
  O.~Aharony and Y.~Tachikawa,
  ``A Holographic computation of the central charges of d=4, N=2 SCFTs,''
  JHEP {\bf 0801}, 037 (2008)
  [arXiv:0711.4532 [hep-th]].
  

  
  










\bibitem{Witten}
  E.~Witten,
  ``Solutions of four-dimensional field theories via M theory,''
  Nucl.\ Phys.\  {\bf B500}, 3-42 (1997).
  [hep-th/9703166].

%
%
%
  


\bibitem{DW}
  R.~Donagi and E.~Witten,
  ``Supersymmetric Yang-Mills Theory And Integrable Systems,''
  Nucl.\ Phys.\  B {\bf 460}, 299 (1996)
  [arXiv:hep-th/9510101].

\bibitem{NX1}
  D.~Nanopoulos, D.~Xie,
  ``Hitchin Equation, Singularity, and N=2 Superconformal Field Theories,''
  JHEP {\bf 1003}, 043 (2010).
  [arXiv:0911.1990 [hep-th]].

\bibitem{Hitchin}
  N.~J.~Hitchin,
  ``Stable bundles and integrable systems,''
  Duke Math.\ J.\  {\bf 54}, 91-114 (1987).
  
\bibitem{Hitchin1}
  N.~J.~Hitchin,
  ``The Selfduality equations on a Riemann surface,''
  Proc.\ Lond.\ Math.\ Soc.\  {\bf 55}, 59-131 (1987).

\bibitem{Simpson}
  C.~Simpson,
  ``Harmonic Bundles On Noncompact Curves,"
  J. Amer. Math. Soc. 3 (1990), 713-770.

\bibitem{Markman}
  E.~Markman,
  ``Spectral curves and integrable systems,"
  Comp.\ Math.\ {\bf 93}, 255-290 (1994).

\bibitem{DonagiMarkman} 
  R.~Donagi and E.~Markman,
  ``Spectral curves, algebraically completely integrable Hamiltonian systems, and moduli of bundles,''
  alg-geom/9507017.

  
\bibitem{BB}
  O.~Biquard, P.~ Boalch,
  ``Wild nonabelian Hodge theory on curves,"
  [math.DG/0111098].

\bibitem{NX2}
  D.~Nanopoulos, D.~Xie,
  ``Hitchin Equation, Irregular Singularity, and $N=2$ Asymptotical Free Theories,''
  [arXiv:1005.1350 [hep-th]].

\bibitem{SW3d}
  N.~Seiberg, E.~Witten,
  ``Gauge dynamics and compactification to three-dimensions,''
  [hep-th/9607163].

\bibitem{HMS}
  J.~A.~Harvey, G.~W.~Moore, A.~Strominger,
  ``Reducing S duality to T duality,''
  Phys.\ Rev.\  {\bf D52}, 7161-7167 (1995).
  [hep-th/9501022].
  
\bibitem{BJSV}
  M.~Bershadsky, A.~Johansen, V.~Sadov, C.~Vafa,
  ``Topological reduction of 4-d SYM to 2-d sigma models,''
  Nucl.\ Phys.\  {\bf B448}, 166-186 (1995).
  [hep-th/9501096].


\bibitem{KW}
  A.~Kapustin, E.~Witten,
  ``Electric-Magnetic Duality And The Geometric Langlands Program,''
  [hep-th/0604151].
  
\bibitem{GW}
  S.~Gukov, E.~Witten,
  ``Gauge Theory, Ramification, And The Geometric Langlands Program,''
  [hep-th/0612073].
  
\bibitem{WildWitten}
  E.~Witten,
  ``Gauge theory and wild ramification,''
  [arXiv:0710.0631 [hep-th]].

\bibitem{CK}
  S.~A.~Cherkis, A.~Kapustin,
  ``Singular monopoles and supersymmetric gauge theories in three-dimensions,''
  Nucl.\ Phys.\  {\bf B525}, 215-234 (1998).
  [hep-th/9711145].

\bibitem{Kapustin}
  A.~Kapustin,
  ``Solution of N=2 gauge theories via compactification to three-dimensions,''
  Nucl.\ Phys.\  {\bf B534}, 531-545 (1998).
  [hep-th/9804069].






  
  
  
  
  
  
  
  
  
  
  
  
  
  
  
  

\bibitem{Taki} 
  M.~Taki,
  ``On AGT Conjecture for Pure Super Yang-Mills and W-algebra,''
  JHEP\ {\bf 1105}, 038  (2011)
  [arXiv:0912.4789 [hep-th]].

\bibitem{KPPW} 
  C.~Kozcaz, S.~Pasquetti, F.~Passerini and N.~Wyllard,
  ``Affine sl(N) conformal blocks from N=2 SU(N) gauge theories,''
  JHEP\ {\bf 1101}, 045  (2011)
  [arXiv:1008.1412 [hep-th]].
  
\bibitem{Wyllard} 
  N.~Wyllard,
  ``W-algebras and surface operators in N=2 gauge theories,''
  J.\ Phys.\ AA\ {\bf 44}, 155401  (2011)
  [arXiv:1011.0289 [hep-th]];
%
  ``Instanton partition functions in N=2 SU(N) gauge theories with a general surface operator, and their W-algebra duals,''
  JHEP\ {\bf 1102}, 114  (2011)
  [arXiv:1012.1355 [hep-th]];\\
%
  H.~Kanno and Y.~Tachikawa,
  ``Instanton counting with a surface operator and the chain-saw quiver,''
  JHEP\ {\bf 1106}, 119  (2011)
  [arXiv:1105.0357 [hep-th]].
  
\bibitem{BF} 
  V.~Belavin and B.~Feigin,
  ``Super Liouville conformal blocks from N=2 SU(2) quiver gauge theories,''
  JHEP\ {\bf 1107}, 079  (2011)
  [arXiv:1105.5800 [hep-th]];\\
%
  G.~Bonelli, K.~Maruyoshi and A.~Tanzini,
  ``Instantons on ALE spaces and Super Liouville Conformal Field Theories,''
  arXiv:1106.2505 [hep-th];\\
%
  Y.~Ito,
  ``Ramond sector of super Liouville theory from instantons on an ALE space,''
  arXiv:1110.2176 [hep-th].

\bibitem{WyllardALE} 
  N.~Wyllard,
  ``Coset conformal blocks and N=2 gauge theories,''
  arXiv:1109.4264 [hep-th].

\bibitem{KMSTach} 
  C.~A.~Keller, N.~Mekareeya, J.~Song and Y.~Tachikawa,
  ``The ABCDEFG of Instantons and W-algebras,''
  arXiv:1111.5624 [hep-th].

\bibitem{BE} 
  A.~Braverman and P.~Etingof,
  ``Instanton counting via affine Lie algebras II: From Whittaker vectors to the Seiberg-Witten prepotential,''
  math/0409441 [math-ag].
  
  
\bibitem{MMM}
  A.~Marshakov, A.~Mironov and A.~Morozov,
  ``On non-conformal limit of the AGT relations,''
  Phys.\ Lett.\  B {\bf 682}, 125 (2009)
  [arXiv:0909.2052 [hep-th]].

\bibitem{HJS}
  L.~Hadasz, Z.~Jaskolski and P.~Suchanek,
  ``Proving the AGT relation for $N_f = 0,1,2$ antifundamentals,''
  JHEP {\bf 1006}, 046 (2010)
  [arXiv:1004.1841 [hep-th]].

\bibitem{Zamolodchikov:1985ie}
  A.~B.~Zamolodchikov,
  ``Conformal Symmetry In Two-Dimensions: An Explicit Recurrence Formula For
  The Conformal Partial Wave Amplitude,''
  Commun.\ Math.\ Phys.\  {\bf 96}, 419 (1984).

\bibitem{Poghossian}
  R.~Poghossian,
  ``Recursion relations in CFT and N=2 SYM theory,''
  JHEP {\bf 0912}, 038 (2009)
  [arXiv:0909.3412 [hep-th]].
  
\bibitem{HJS0}
  L.~Hadasz, Z.~Jaskolski and P.~Suchanek,
  ``Recursive representation of the torus 1-point conformal block,''
  arXiv:0911.2353 [hep-th].

\bibitem{FL}
  V.~A.~Fateev and A.~V.~Litvinov,
  ``On AGT conjecture,''
  JHEP {\bf 1002}, 014 (2010)
  [arXiv:0912.0504 [hep-th]].

\bibitem{Matone}
  M.~Matone,
  ``Instantons and recursion relations in N=2 SUSY gauge theory,''
  Phys.\ Lett.\  B {\bf 357}, 342 (1995)
  [arXiv:hep-th/9506102].









\bibitem{BMTY} 
  G.~Bonelli, K.~Maruyoshi, A.~Tanzini and F.~Yagi,
  ``Generalized matrix models and AGT correspondence at all genera,''
  JHEP\ {\bf 1107}, 055  (2011)
  [arXiv:1011.5417 [hep-th]].

\bibitem{Hollands}
  L.~Hollands, C.~A.~Keller, J.~Song,
  ``Towards a 4d/2d correspondence for Sicilian quivers,''
  [arXiv:1107.0973 [hep-th]].






\bibitem{BPZ}
  A.~A.~Belavin, A.~M.~Polyakov, A.~B.~Zamolodchikov,
  ``Infinite Conformal Symmetry in Two-Dimensional Quantum Field Theory,''
  Nucl.\ Phys.\  {\bf B241}, 333-380 (1984).

\bibitem{AY}
  H.~Awata, Y.~Yamada,
  ``Five-dimensional AGT Conjecture and the Deformed Virasoro 'lgebra,''
  JHEP {\bf 1001}, 125 (2010).
  [arXiv:0910.4431 [hep-th]].
  
\bibitem{MT} 
  K.~Maruyoshi and M.~Taki,
  ``Deformed Prepotential, Quantum Integrable System and Liouville Field Theory,''
  Nucl.\ Phys.\ B\ {\bf 841}, 388  (2010)
  [arXiv:1006.4505 [hep-th]].

\bibitem{Awata:2010bz}
  H.~Awata, H.~Fuji, H.~Kanno, M.~Manabe, Y.~Yamada,
  ``Localization with a Surface Operator, Irregular Conformal Blocks and Open Topological String,''
  [arXiv:1008.0574 [hep-th]].

\bibitem{AGGTV}
  L.~F.~Alday, D.~Gaiotto, S.~Gukov, Y.~Tachikawa, H.~Verlinde,
  ``Loop and surface operators in N=2 gauge theory and Liouville modular geometry,''
  JHEP {\bf 1001}, 113 (2010).
  [arXiv:0909.0945 [hep-th]].
  
\bibitem{KPW}
  C.~Kozcaz, S.~Pasquetti, N.~Wyllard,
  ``A \& B model approaches to surface operators and Toda theories,''
  JHEP {\bf 1008}, 042 (2010).
  [arXiv:1004.2025 [hep-th]].
  
\bibitem{DGH}
  T.~Dimofte, S.~Gukov, L.~Hollands,
  ``Vortex Counting and Lagrangian 3-manifolds,''
  [arXiv:1006.0977 [hep-th]].
  
\bibitem{BTZ}
  G.~Bonelli, A.~Tanzini, J.~Zhao,
  ``The Liouville side of the Vortex,''
  JHEP {\bf 1109}, 096 (2011).
  [arXiv:1107.2787 [hep-th]].
  

\bibitem{Teschner} 
  J.~Teschner,
  ``Quantization of the Hitchin moduli spaces, Liouville theory, and the geometric Langlands correspondence I,''
  arXiv:1005.2846 [hep-th].

\bibitem{MMMint} 
  A.~Marshakov, A.~Mironov and A.~Morozov,
  ``On AGT Relations with Surface Operator Insertion and Stationary Limit of Beta-Ensembles,''
  J.\ Geom.\ Phys.\ \ {\bf 61}, 1203  (2011)
  [arXiv:1011.4491 [hep-th]].

\bibitem{BMTintegral} 
  G.~Bonelli, K.~Maruyoshi and A.~Tanzini,
  ``Quantum Hitchin Systems via beta-deformed Matrix Models,''
  arXiv:1104.4016 [hep-th].

\bibitem{Tai} 
  T.~-S.~Tai,
  ``Uniformization, Calogero-Moser/Heun duality and Sutherland/bubbling pants,''
  JHEP\ {\bf 1010}, 107  (2010)
  [arXiv:1008.4332 [hep-th]].

\bibitem{Piatek} 
  M.~Piatek,
  ``Classical conformal blocks from TBA for the elliptic Calogero-Moser system,''
  JHEP\ {\bf 1106}, 050  (2011)
  [arXiv:1102.5403 [hep-th]].
  
\bibitem{FFR} 
  B.~Feigin, E.~Frenkel and N.~Reshetikhin,
  ``Gaudin model, Bethe ansatz and correlation functions at the critical level,''
  Commun.\ Math.\ Phys.\ \ {\bf 166}, 27  (1994)
  [hep-th/9402022].
  
\bibitem{Frenkel} 
  E.~Frenkel,
  ``Gaudin model and opers,''
  math/0407524 [math-qa].

\bibitem{FFT} 
  B.~Feigin, E.~Frenkel and V.~Toledano Laredo,
  ``Gaudin models with irregular singularities,''
  Adv.\ Math.\ \ {\bf 223}, 873  (2010)
  [math/0612798 [math.QA]].
  
  
  






  
  

\bibitem{IMO} 
  H.~Itoyama, K.~Maruyoshi and T.~Oota,
  ``The Quiver Matrix Model and 2d-4d Conformal Connection,''
  Prog.\ Theor.\ Phys.\ \ {\bf 123}, 957  (2010)
  [arXiv:0911.4244 [hep-th]].

\bibitem{KMST} 
  S.~Kanno, Y.~Matsuo, S.~Shiba and Y.~Tachikawa,
  ``N=2 gauge theories and degenerate fields of Toda theory,''
  Phys.\ Rev.\ D\ {\bf 81}, 046004  (2010)
  [arXiv:0911.4787 [hep-th]].
  
\bibitem{ND3} 
  D.~Nanopoulos and D.~Xie,
  ``$N=2$ Generalized Superconformal Quiver Gauge Theory,''
  arXiv:1006.3486 [hep-th].
  
\bibitem{CD} 
  O.~Chacaltana and J.~Distler,
  ``Tinkertoys for Gaiotto Duality,''
  JHEP\ {\bf 1011}, 099  (2010)
  [arXiv:1008.5203 [hep-th]].

\bibitem{DP} 
  N.~Drukker and F.~Passerini,
  ``(de)Tails of Toda CFT,''
  JHEP\ {\bf 1104}, 106  (2011)
  [arXiv:1012.1352 [hep-th]].

\bibitem{Tachikawa} 
  Y.~Tachikawa,
  ``On W-algebras and the symmetries of defects of 6d N=(2,0) theory,''
  JHEP\ {\bf 1103}, 043  (2011)
  [arXiv:1102.0076 [hep-th]].



\bibitem{DV}
  R.~Dijkgraaf, C.~Vafa,
  ``Toda Theories, Matrix Models, Topological Strings, and N=2 Gauge Systems,''
  [arXiv:0909.2453 [hep-th]].
  
\bibitem{EM} 
  T.~Eguchi and K.~Maruyoshi,
  ``Penner Type Matrix Model and Seiberg-Witten Theory,''
  JHEP\ {\bf 1002}, 022  (2010)
  [arXiv:0911.4797 [hep-th]];
  %
  ``Seiberg-Witten theory, matrix model and AGT relation,''
  JHEP\ {\bf 1007}, 081  (2010)
  [arXiv:1006.0828 [hep-th]].
  
\bibitem{MMS} 
  A.~Mironov, A.~Morozov and S.~.Shakirov,
  ``Brezin-Gross-Witten model as 'pure gauge' limit of Selberg integrals,''
  JHEP\ {\bf 1103}, 102  (2011)
  [arXiv:1011.3481 [hep-th]].
  

\bibitem{NS1} 
  N.~A.~Nekrasov and S.~L.~Shatashvili,
  ``Quantization of Integrable Systems and Four Dimensional Gauge Theories,''
  arXiv:0908.4052 [hep-th].
  
\bibitem{NS2} 
  N.~Nekrasov, A.~Rosly and S.~Shatashvili,
  ``Darboux coordinates, Yang-Yang functional, and gauge theory,''
  Nucl.\ Phys.\ Proc.\ Suppl.\ \ {\bf 216}, 69  (2011)
  [arXiv:1103.3919 [hep-th]].
  

\bibitem{HKK} 
  M.~-x.~Huang, A.~-K.~Kashani-Poor and A.~Klemm,
  ``The Omega deformed B-model for rigid N=2 theories,''
  arXiv:1109.5728 [hep-th].

\end{thebibliography}
\end{document}